\documentclass[11pt]{article}
\pdfoutput=1
\usepackage{jheppub}
\usepackage{amsmath,amssymb,mathtools,bm}
\usepackage{booktabs}
\usepackage{array}
\usepackage{xcolor}
\usepackage{hyperref}
\hypersetup{colorlinks=true,linkcolor=blue!55!black,citecolor=blue!55!black,urlcolor=blue!55!black}

\newcommand{\dd}{\mathrm d}
\newcommand{\Tr}{\operatorname{Tr}}
\newcommand{\Vol}{\operatorname{Vol}}

\newcommand{\cR}{\mathcal R}
\newcommand{\cP}{\mathcal P}
\newcommand{\cA}{\mathcal A}

\newcommand{\cJ}{\mathcal J}

\newcommand{\St}{\mathfrak S}

\begin{document}

\title{Hironaka Geometry and Resurgence in Finite-$N$ Matrix Models}
\author[a,b]{Robert de Mello Koch,}
\author[a]{Vinayak Raj}
\author[a,b]{and Anik Rudra}
\affiliation[a]{School of Science, Huzhou Normal University, Huzhou 313000, China}
\affiliation[b]{Mandelstam Institute for Theoretical Physics, School of Physics, University of the Witwatersrand, Private Bag 3, Wits 2050, South Africa}
\emailAdd{robert@zjhu.edu.cn}
\emailAdd{vinayak.hep.th@gmail.com}
\emailAdd{anikrudra23@gmail.com}
\date{September 2026}

\abstract{We study how finite-$N$ invariant theory organizes the non-perturbative structure of matrix models. For a model of four traceless Hermitian $2\times 2$ matrices, the Hironaka decomposition realizes the gauge-invariant configuration space as an eight-sheeted branched cover of the space of primary invariants. We show that ramification points of this cover naturally generate additional saddle points when the action depends only on the primaries. For an explicit rank-two saddle, we compute its action, one-loop normalization and Picard--Lefschetz connection to the perturbative vacuum. The same saddle controls the leading Borel singularity and large-order growth of perturbation theory, while its first fluctuation correction reproduces the first subleading large-order correction with no fitted parameters. Our results provide a concrete link between finite-$N$ invariant geometry and resurgence in matrix models.}

\maketitle

\section{Introduction}

The AdS/CFT correspondence~\cite{Maldacena:1997re,Gubser:1998bc,Witten:1998qj} provides a non-perturbative formulation of quantum gravity in terms of an ordinary conformal field theory living in one fewer dimensions. In the holographic dictionary, the bulk gravitational coupling is controlled parametrically by $1/N^2$, so that the classical gravity limit corresponds to $N\to\infty$. Finite-$N$ effects in the CFT therefore encode quantum-gravitational corrections, with effects that are non-perturbative in $1/N$ providing a natural window into genuinely non-perturbative aspects of quantum gravity. In this paper, we take a step toward understanding such non-perturbative quantum-gravitational effects by studying their finite-$N$ manifestation in CFTs with $N\times N$ matrix degrees of freedom.

A prototypical example of the class of theories we have in mind is $\mathcal{N}=4$ super Yang--Mills theory. This theory admits several interesting limits and truncations in which the dynamics reduces to an ordinary quantum-mechanical system. Two particularly clean examples are the BMN matrix model~\cite{Berenstein:2002jq} and spin matrix theory~\cite{Harmark:2014mpa,Baiguera:2022pll}. It is therefore natural to ask how effects that are non-perturbative in $1/N$ manifest themselves in matrix quantum mechanics. In this article, we take some steps towards answering this question.

An important source of finite-$N$ effects is provided by trace relations. Gauge-invariants are generated by the cyclically distinct single-trace operators, with general gauge-invariant operators obtained as polynomials in these generators. At infinite $N$, these polynomials are independent, yielding a complete and non-redundant description of the gauge-invariant operator space. At finite $N$ however, trace relations introduce nontrivial relations~\cite{Pr,Razmyslov,DrenskyFormanek} among operators independent at large-$N$, so that the unrestricted set of polynomials in single traces becomes overcomplete. Constructing a complete and non-redundant basis therefore requires solving the infinite set of trace relations among the infinite collection of gauge-invariant operators. This problem was recently addressed systematically in~\cite{deMelloKoch:2025ngs,deMelloKoch:2025rkw,deMelloKochKimVanZyl:2025,deMelloKochRodrigues:2026}.

It turns out that even at finite $N$ one can still describe the complete set of gauge invariant operators using suitable generators~\cite{deMelloKoch:2025ngs}. The finite $N$ problem has a richer structure and the generators can be classified as one of two types: either primary (denoted $p_i$) or secondary (denoted $\eta_\alpha$) invariants. Gauge invariant operators are still polynomials in these invariants, but with some extra rules. One can freely use the primary invariants: any monomial can include a product of any number of different primary invariants, each raised to any non-negative power. The number of primary invariants grows as $1+(d-1)N^2$. On the other hand, each monomial must include a single secondary invariant and this must appear linearly. The number of secondary invariants, which grows as~\cite{deMelloKochKimVanZyl:2025} $e^{cN^2}$ with $c$ an $O(1)$ number, has the parametric growth required to accommodate the entropy of a black hole. For complete details we refer the reader to~\cite{deMelloKoch:2025ngs,OurReview}. The upshot is that the gauge invariant operators can be organized into a complete and non-redundant set $GIO$ in terms of the Hironaka decomposition
\begin{equation}
GIO=\left\{\bigoplus_{\alpha,\{n_i\}} \prod_i (p_i)^{n_i}\eta_\alpha\right\}
\end{equation}
where each term in the direct sum includes a single secondary invariant dressed by arbitrary powers of the primary invariants. Mathematically GIO is a free module, with basis given by the secondary invariants, defined over the free ring generated by the primary invariants.

Although the decomposition of the algebra of invariant operators is, at first sight, a rather abstract algebraic construction, its physical meaning becomes clearer when one recognizes that it also induces a decomposition of the Hilbert space of the matrix oscillator~\cite{OurReview}. In this interpretation the primary invariants are promoted to Fock-space creation operators, while the secondary invariants organize finite sectors, each of which supports an entire Fock-space tower generated by the primaries. This structure is highly suggestive of an organization of the Hilbert space into distinct non-perturbative sectors, with each sector supporting its own perturbative tower of excitations. See~\cite{OurReview} for a detailed discussion.

The Hironaka decomposition also admits a natural physical interpretation from the path-integral point of view~\cite{deMelloKochRodrigues:2026,deMelloKoch:2026pck}. The cleanest setting in which to see this is provided by finite-dimensional matrix integrals. In this case, one may change variables from the matrix elements appearing in the original integral to a set of invariant variables built from the primary invariants. The full invariant space is then realized as a finite-sheeted cover of the space coordinatized by the primaries, with the different sheets distinguished by the secondary invariants. The number of sheets matches the number of secondary invariants. From this perspective, the Hironaka decomposition provides a \emph{global algebraic} description of the invariant configuration space as a finite-sheeted cover of the primary-invariant space.

There is a second reason why the invariant formulation is natural for the questions that motivate us. Gauge-invariant variables are precisely the variables adapted to the large-$N$ expansion~\cite{Jevicki:1979mb,Jevicki:1980zg,Das:1990kaa}. Their normalization must be chosen so that they remain finite as $N\to\infty$. For example, if the matrix elements are kept fixed as $N$ grows, one generically has ${\rm Tr} (X^{2n})\sim N^{n+1}$; equivalently, after writing  $\hat{X}=X/\sqrt{N}$, the normalized moments
\begin{equation}
\phi_k\,\,=\,\, \frac1N {\rm Tr}\hat{X}^k
\end{equation}
are $O(1)$ at large $N$. When the matrix integral is rewritten in terms of such normalized invariant variables, the action together with the quotient Jacobian organizes itself into a large-$N$ effective action of the schematic form
\begin{equation}
Z=\int [d\phi ]\exp[-N^2 S_{\rm eff}(\phi)].
\end{equation}
Expanding about a saddle, $\phi=\phi_\star+N^{-1}\eta$, then produces a conventional loop expansion, with interaction vertices suppressed by powers of $1/N$. Thus the invariant-variable description is not only a convenient way of implementing finite-$N$ trace relations; it is also the natural starting point for the semiclassical $1/N$ expansion.

In the explicit model studied below we set $N=2$. This is essential for obtaining a nontrivial Hironaka cover whose algebraic geometry and resurgent structure can both be analyzed essentially completely, but it removes the explicit large-$N$ bookkeeping from the problem. The expansion parameter used in our concrete calculation is therefore the coupling $g$, rather than $1/N$. Nevertheless, the invariant variables in which the calculation is formulated are precisely those which, for general $N$ and after the appropriate normalization, furnish the natural variables for the $1/N$ expansion. The $N=2$ example should therefore be viewed as an exactly tractable finite-$N$ laboratory for a geometric structure that is naturally embedded in large-$N$ perturbation theory.

We now turn to what may at first appear to be a completely different subject: perturbative expansions in quantum mechanics and quantum field theory. To construct such an expansion, one identifies a small parameter, such as a coupling constant $g$ or $\hbar$, and expresses the observable of interest as a formal power series in this parameter. Generically however, the resulting perturbative series is not convergent but only asymptotic. Consider, for example,
\begin{equation}
{\cal O}(g)= \sum_{n=0}^\infty c_n g^n ,
\end{equation}
with coefficients exhibiting factorial growth $c_n\sim n!$. A standard way to extract information from such a divergent series is Borel resummation (see~\cite{Aniceto:2011nu,Marino:2012zq,Dorigoni:2014hea,ABS,Dunne:2014bca} and references therein). One first defines the Borel transform
\begin{equation}
{\cal B}[{\cal O}](x)=\sum_{n=0}^\infty \frac{c_n}{n!}x^n .
\end{equation}
The factorial growth of the original coefficients has now been tamed, and the Borel transform typically has a nonzero radius of convergence. It can then be analytically continued beyond this radius, and the original observable is formally reconstructed by a Laplace transform,
\begin{equation}
{\cal S}{\cal O}(g)=\frac{1}{g}\int_0^\infty dx\,e^{-x/g}\,{\cal B}[{\cal O}](x).
\end{equation}
The divergence of the original perturbative expansion reappears in a new guise: the analytically continued Borel transform develops singularities in the Borel plane. Remarkably, the locations and structure of these singularities are tied to non-perturbative saddle points of the path integral, including instanton and multi-instanton sectors. Resurgence makes this relation precise by organizing the perturbative expansions about the different perturbative and non-perturbative saddles into a single transseries. In this way, information contained in the divergent perturbative coefficients can be combined with the contributions of non-perturbative sectors to reconstruct the \emph{global analytic} structure of the observable.

The Hironaka decomposition encodes \emph{global algebraic information} about the finite-$N$ invariant ring, while resurgence reconstructs a \emph{global analytic answer} from perturbative data defined locally around a given saddle. It is therefore natural to ask how much of the global structure uncovered by resurgence is already encoded in the Hironaka decomposition. There is, however, an important distinction between the two frameworks. The Hironaka decomposition is entirely kinematical: it follows from the field content and the action of the gauge symmetry, without making any reference to the dynamics or to a particular choice of action. Resurgence, by contrast, is intrinsically dynamical and depends crucially on the structure of the action and its saddle points. Nevertheless, because both frameworks capture global information that is invisible in a purely local description, one may expect a nontrivial relation between them. We will see that an interesting overlap does indeed arise.

We study a matrix model built from four traceless Hermitian $2\times 2$ matrices $X^a$. In Section~\ref{HDecomp}, we construct the Hironaka decomposition of the corresponding invariant ring and show that, at finite $N$, the invariant space takes the form of a finite branched cover of the space coordinatized by the primary invariants. For the model considered here, this algebraic cover has eight sheets. Our basic observable is the finite-dimensional matrix integral
\begin{equation}
Z(g)=\int \prod_{a=1}^4 dX^a e^{-\frac{V(X^a)}{g}},
\end{equation}
with a quartic potential. As emphasized particularly clearly in the review~\cite{ABS}, finite-dimensional integrals provide useful laboratories in which the essential ingredients of resurgence are already present.

The integral above is initially a twelve-dimensional integral over the matrix elements of the four traceless matrices $X^a$. Quotienting by the three-dimensional gauge symmetry leaves nine independent gauge-invariant degrees of freedom. In Section~\ref{JacobQuotient}, we change variables from the original matrix elements to the nine primary invariants and determine the associated Jacobian. A perturbative expansion about a chosen vacuum probes a local neighborhood on a single branch of the eight-sheeted cover. In Section~\ref{PT}, we explain how to generate this perturbative expansion explicitly; the resulting construction can be implemented straightforwardly in Mathematica and pushed to high orders; we will work to the 50th order. Working directly with the integral expressed in terms of the primary invariants, we are furthermore able to obtain an exact angular-integral representation for the Borel--Leroy transform of the perturbative series. This result, derived in Section~\ref{BorelLeroy}, makes it possible to determine the singularity structure of the Borel transform analytically.

These singularities admit a particularly simple interpretation in terms of the geometry of the Hironaka cover. We find that ramification loci provide the natural geometric origin of candidate non-perturbative saddles, and when the remaining dynamical and Picard–Lefschetz conditions are satisfied, these saddles generate Borel singularities. This correspondence is developed in detail in Section~\ref{Ram}. In this way, the ramification geometry of the finite-$N$ invariant space becomes directly visible to perturbation theory through resurgence.

More concretely, we identify a point at which two sheets of the Hironaka cover collide, show that this point corresponds to a genuine saddle of the original twelve-dimensional matrix integral, and compute both its action and its one-loop determinant. The same saddle action is then recovered from the large-order growth of perturbation theory performed about a vacuum lying on a different sheet. We thereby obtain a concrete chain of relations linking the finite-$N$ invariant algebra to the branched cover of primary space, its ramification loci, the associated additional saddles of the matrix integral, and finally the singularities of the Borel transform. This provides a direct bridge between finite-$N$ invariant theory and non-perturbative physics.

Resurgence already teaches us that perturbation theory about one saddle can encode information about other saddles and, more generally, about the global structure of the relevant Lefschetz thimbles. The new ingredient in the present work is the observation that finite-$N$ invariant theory supplies an independent algebraic structure that organizes these non-perturbative saddles. In our example, the Hironaka decomposition does not merely provide a convenient parametrization of the invariant ring: the geometry of its finite branched cover is directly reflected in the resurgent structure of perturbation theory.

A few comments are in order. Although the number of secondary invariants equals the number of sheets of the finite cover, the secondary invariants themselves should not be identified with individual sheets. There is, in general, no canonical correspondence $\eta_\alpha\leftrightarrow$sheet$_\alpha$, and any physical interpretation assigned to a particular choice\footnote{Choosing a set of secondary invariants corresponds to choosing a module basis.} of Hironaka secondaries is basis dependent. Rather, the secondaries furnish a basis for the finite fiber algebra, while the individual sheets are naturally singled out by the primitive idempotents of this algebra. For the example studied here, as explained in Section~\ref{HDecomp}, the generic fiber algebra admits the presentation
\begin{equation}
\cA_{\mathbb K}=\mathbb K[\Sigma,\Omega]/(F,\Omega^2-D),
\end{equation}
The eight generic points of the fiber are obtained by taking the four roots $\Sigma_\alpha$ of $F$, together with the two possible signs $\Omega_\alpha=\sqrt{D(\Sigma_\alpha)}$. The corresponding primitive idempotents can be constructed explicitly by Lagrange interpolation. Schematically,
\begin{equation}
e_{\alpha,\epsilon}=\left[\prod_{\beta\ne\alpha}\frac{\Sigma-\Sigma_\beta}{\Sigma_\alpha-\Sigma_\beta}\right]\frac12\left(1+\epsilon\frac{\Omega}{\sqrt{D(\Sigma_\alpha)}}\right),\qquad\epsilon=\pm 1.
\end{equation}
These, rather than the individual Hironaka secondaries, are the algebraic objects that localize onto a single sheet of the cover.

A second observation is particularly important for the physical interpretation. The action considered below depends only on the primary invariants,
\begin{equation}
V(X)=V(p(X))=\frac{\lambda}{2}\sum_i p_i(X)^2.
\end{equation}
Consequently,
\begin{equation}
dV_X=(d\pi_X)^T\,dV_p.
\end{equation}
At a critical point of the primary-space potential, and in particular at $p_i=0$, one has $dV_p=0$. It follows immediately that every point in the fiber $\pi^{-1}(0)$ is a critical point of the original complexified matrix integral. In our example this fiber consists of
\begin{equation}
(\Sigma,\Omega)=(\pm1,\pm1),\qquad (\pm\sqrt3,\pm i).
\end{equation}
Thus the finite Hironaka fiber above the perturbative primary vacuum already contains eight algebraically distinct critical configurations in the complexified quotient. Four lie on the real positive-semidefinite slice, while four are intrinsically complex. In this precise sense, for an action depending only on the primaries, the finite secondary fiber over a primary critical point organizes a finite collection of distinct saddle sectors.

There is, however, an important qualification. Because the action depends only on the primaries, all points lying in the same fiber have exactly the same action. Distinct sheets are therefore not, by themselves, separated by non-perturbative factors of the form $e^{-A/g}$. The fiber points instead represent distinct saddle or vacuum sectors that are invisible to the primary action. The genuinely non-perturbative scales arise from additional saddles associated with the geometry connecting these sectors, and in particular with ramification loci where sheets collide. These ramification saddles carry nonzero action differences and are detected as singularities of the Borel transform. The role of the secondary algebra is therefore not simply to enumerate non-perturbative exponentials, but to organize the finite set of saddle sectors and the algebraic geometry of their collisions.

Finally, we note that this work goes substantially further than the study~\cite{deMelloKochRodrigues:2026} which established the finite-cover and sector interpretation, in the context of a finite matrix integral. The new result here is the connection of the ramification geometry to genuine saddles and resurgence, including the full-matrix Picard–Lefschetz incidence, the one-loop coefficient, and a parameter-free first subleading bridge-equation test.

\section{Invariant theory and the Hilbert series}\label{HDecomp}

As was mentioned in the introduction, we study the matrix model of $d=4$ traceless Hermitian $2\times2$ matrices, which can be expanded as
\begin{equation}
 X^a=\vec x_a\cdot\vec\sigma, \qquad a=1,\ldots,4,
\end{equation}
where $\vec x_a\in\mathbb R^3$ and $\vec\sigma$ are the Pauli matrices.  In terms of the original matrix variables, the model has a $U(2)$ gauge symmetry which acts as $X^a\to UX^a U^\dagger$, $U\in U(2)$. In terms of the vectors $\vec x_a$, the gauge symmetry acts as an overall rotation $\vec x_a\to R(U)\vec x_a$ where $R(U)\in SO(3)$ is a rotation determined by $U$. Since
\begin{equation}
 \frac12\Tr(X^aX^b)=\vec x_a\cdot\vec x_b,
\end{equation}
all degree 2 gauge-invariants are encoded in the Gram matrix $G_{ab}=\vec x_a\cdot\vec x_b$. Since our four vectors live in three dimensions, they must obey a relation of the form
\begin{equation}
c_1\vec x_1+c_2\vec x_2+c_3\vec x_3+c_4\vec x_4=0.
\end{equation}
Taking a dot product of this equation with $\vec{x}_a$ and writing the result in terms of the Gram matrix we find
\begin{equation}
c_1 G_{1a}+c_2 G_{2a}+c_3 G_{3a}+c_4 G_{4a}=0
\end{equation}
which implies that the rows of the Gram matrix are never linearly independent and hence
\begin{equation}
 \det G=0.
\end{equation}
Since the vectors $\vec{x}_a$ are three-dimensional it is easy to generalize the above argument to conclude that generically\footnote{Generically the four vectors do not lie in a two-dimensional plane.} $G$ has rank 3. Thus ten symmetric Gram entries are subject to one relation, leaving nine continuous invariant directions, which agrees with the elementary dimension count of degrees of freedom minus gauge parameters
\begin{equation}
 4\times3-\dim SO(3)=12-3=9.
\end{equation}
We use the following nine primary invariants
\begin{align}
 p_1&=G_{11}-1,& p_2&=G_{33}-1,\nonumber\\
 p_3&=G_{24}-1-\frac14(G_{22}+G_{44}-4),&
 p_4&=G_{22}+G_{44}-4,\nonumber\\
 p_5&=G_{22}-G_{44},& p_6&=G_{12}-G_{34},\nonumber\\
 p_7&=G_{13},&p_8&=G_{14},&p_9&=G_{23}.\label{eq:udef}
\end{align}
The remaining quadratic Gram coordinate is $\Sigma=G_{12}$ and we can write $G_{34}=\Sigma-p_6$. Therefore
\begin{equation}
G(\vec p,\Sigma)=
\begin{pmatrix}
1+p_1&\Sigma&p_7&p_8\\
\Sigma&2+\frac{p_4+p_5}{2}&p_9&1+p_3+\frac{p_4}{4}\\
p_7&p_9&1+p_2&\Sigma-p_6\\
p_8&1+p_3+\frac{p_4}{4}&\Sigma-p_6&2+\frac{p_4-p_5}{2}
\end{pmatrix}, \label{eq:G}
\end{equation}
and the rank-three condition is $F(\vec p,\Sigma)=\det G(\vec p,\Sigma)=0$. We choose a simple potential
\begin{equation}
 V(\vec p)=\frac{\lambda}{2}\sum_{i=1}^9p_i^2,\label{eq:Vsimple}\qquad \qquad H_{ij}=\frac{\partial^2V}{\partial p_i\partial p_j} =\lambda\delta_{ij}.
\end{equation}
Every primary direction therefore has the same stiffness.

We are considering the ring of invariants of $d=4$, $N\times N$ matrices, invariant under conjugation by $U(N)$, with $N=2$. A straightforward computation gives the Hilbert series of this ring (see for example \cite{OurReview})
\begin{equation}
 \mathcal H(t)= \frac{1+t^2+4t^3+t^4+t^6}{(1-t)^4(1-t^2)^9}. \label{eq:Hilbert-full}
\end{equation}
The denominator has 13 factors, which matches the Krull dimension. We have restricted to traceless matrices which removes the 4 degree one invariants and so for our traceless matrices we should use
\begin{equation}
 \mathcal H(t)= \frac{1+t^2+4t^3+t^4+t^6}{(1-t^2)^9}. \label{eq:Hilbert-traceless}
\end{equation}
We read off 9 primary invariants which are all of degree 2, from the denominator, and 8 secondary invariants from the numerator. Although several of the $p_i$ in \eqref{eq:udef} have been shifted by constants so that a chosen vacuum sits at $\vec p=0$, they generate the same polynomial subring as the corresponding homogeneous quadratic invariants; the degree assignments below refer to those homogeneous generators. The primary invariants freely generate the polynomial subring
\begin{equation}
 \cP=\mathbb C[p_1,\ldots,p_9],\qquad \deg p_i=2,
\end{equation}
and our ring of gauge invariant operators is a free module over this subring, with 8 module generators given by the secondary invariants. The numerator tells us the degrees of a module basis.  We can choose secondaries with degrees
\begin{equation}
 0,\quad 2,\quad 3,3,3,3,\quad 4,\quad 6.
\end{equation}
Thus, our ring of gauge invariant operators can be written as
\begin{equation}
\cR\cong\eta_0\cP\oplus\eta_2\cP\oplus\bigoplus_{a=1}^4\eta_{3,a}\cP \oplus\eta_4\cP\oplus\eta_6\cP \label{eq:Hironaka}
\end{equation}
where we have labelled the secondary invariants $\eta_d$ by their degree $d$. A convenient quadratic secondary is $\eta_2=\Sigma$. Four of the secondary invariants are now given by
\begin{equation}
\eta_0=1,\qquad\eta_2=\Sigma,\qquad\eta_4=\Sigma^2,\qquad\eta_6=\Sigma^3.
\end{equation}
An independent invariant that can be constructed from the vectors $\vec{x}_a$ is given by the cubic orientation secondary
\begin{equation}
T_{abc}=\det(\vec x_a,\vec x_b,\vec x_c)=\vec x_a\cdot (\vec x_b\times \vec x_c).
\end{equation}
In terms of the cubic orientation invariants, the remaining secondary invariants are given by
\begin{equation}
\eta_{3,1}=T_{123},\qquad\quad \eta_{3,2}=T_{124},\qquad\quad \eta_{3,3}=T_{134},\qquad\quad \eta_{3,4}=T_{234}.
\end{equation}
One may verify that these eight invariants are linearly independent over $\mathbb C (p_1,\cdots,p_9)$; comparison with the Hilbert series then shows that they form a Hironaka basis.

The number of secondary invariants is the generic degree of the finite map from invariant space to primary space.  Thus a generic point in primary space has 8 points above it. This is the first appearance of the 8 Hironaka sheets. In general the number of Hironaka sheets is the number of secondary invariants. To develop this idea, a convenient\footnote{We are working on the generic patch where $\vec{x}_1$, $\vec{x}_2$ and $\vec{x}_3$ do not all lie in one plane. In this case the cubic secondary $T_{123}$ is non-vanishing.} cubic secondary is $\Omega=T_{123}$. $\Sigma$ and $\Omega$ obey
\begin{equation}
 F(\vec{p},\Sigma)=0, \qquad \Omega^2=D(\vec{p},\Sigma), \qquad D(\vec{p},\Sigma)=\det G_{123}(\vec{p},\Sigma).
\end{equation}
Here $G_{123}(\vec{p},\Sigma)$ is the $3\times 3$ matrix obtained by restricting to the first three columns and rows of the Gram matrix. Imagine that we choose definite values for all of the 9 primary invariants. In view of \eqref{eq:G} it is clear that $F=\det G$ is quartic in $\Sigma$ so that we get four possible values for $\Sigma$. Further, $\Omega$ is given by a square root of $D$ so we have two possible signs for $\Omega$. Consequently, even after we have chosen definite values for the primary invariants, we have $4\times2=8$ possible values for $\Sigma,\Omega$, i.e. $\Sigma,\Omega$ define an 8-sheeted cover of the primary space. Over the fraction field\footnote{The ring $\cal P$ is not a field because general polynomials don't have an inverse. We can turn $\cal P$ into a field by considering the set of all rational functions of the primary invariants. This is what the fraction field is.} $\mathbb K=\mathbb C(p_1,\ldots,p_9)$, we represent the generic finite algebra\footnote{The generic finite algebra describes the finite set of secondary possibilities lying above a generic choice of primary invariants. It is a finite algebra over the field $\mathbb K$.} as
\begin{equation}
 \cA_{\mathbb K}\simeq \mathbb K[\Sigma,\Omega]/(F,\Omega^2-D).
\end{equation}

For example, at $p_a=0$, for $a=1,2,\cdots,9$, we find
\begin{equation}
 F(\vec 0,\Sigma)=(\Sigma^2-1)(\Sigma^2-3), \qquad D(\vec 0,\Sigma)=2-\Sigma^2.
\end{equation}
Hence the eight algebraic points are
\begin{equation}
 (\Sigma,\Omega)=(\pm1,\pm1), \qquad (\Sigma,\Omega)=(\pm\sqrt3,\pm i).\label{examplefiber}
\end{equation}
Only $(\pm1,\pm1)$ belong to the real positive-semidefinite Hermitian slice, but all eight exist in the complex cover.

\section{The quotient Jacobian}\label{JacobQuotient}

Our next goal is to rewrite the matrix integral in terms of invariant variables. Counting the original number of integration variables, there are 4 vectors each of which is 3 dimensional, giving a total of 12 integration variables. Three of these variables are associated to the $SO(3)$ gauge symmetry, so that only 9 variables remain after we have divided the gauge group out. These 9 variables are the primary invariants. In this section we want to perform the change to invariant variables, divide the gauge group out and derive the resulting invariant measure.

Consider the $3\times 3$ matrix $Y$ produced by taking $\vec{x}_1$, $\vec{x}_2$ and $\vec{x}_3$ as columns
\begin{equation}
 Y=(\vec x_1,\vec x_2,\vec x_3),\qquad q=Y^TY.
\end{equation}
$Y$ includes a total of 9 parameters. We can completely use up the $SO(3)$ gauge symmetry to fix
\begin{equation}
\vec{x}_1=(r_1,0,0)\qquad\vec{x}_2=(a,r_2,0)\qquad \vec{x}_3=(b,c,s r_3)
\end{equation}
where $r_1,r_2,r_3\ge 0$ and $s=\pm 1$. We can't change the sign of the third component of $\vec{x}_3$ while leaving all other components of all three vectors invariant. Choosing the sign of $s$ corresponds to choosing whether the triple $\vec{x}_1,\vec{x}_2,\vec{x}_3$ are right- or left-handed i.e. this is the same sign ambiguity that is present in the cubic invariants. Thus we have two independent $SO(3)$ orbits labelled by the sign of $s$. This argument shows that we can write
\begin{equation}
Y= R T,\qquad T=\left[\begin{matrix} r_1 &a &b\\ 0 &r_2 &c\\ 0 &0 &s r_3\end{matrix}\right],\qquad R\in SO(3).
\end{equation}
This trades the 9 parameters in $Y$ for the 3 parameters of the $SO(3)$ gauge symmetry (in $R$), plus another 6 variables given by $\{r_1,r_2,r_3,a,b,c\}$. Computing the Jacobian as usual we have
\begin{equation}
 \dd^9Y=\sum_{s=\pm 1}r_1^2r_2\,\dd\mu(R)\,\dd r_1\, \dd r_2\, \dd r_3\, \dd a\,\dd b\,\dd c
\end{equation}
Next, notice that
\begin{equation}
(Y^TY)_{ij}=\vec{x}_i\cdot\vec{x}_j=(T^T R^T RT)_{ij}=(T^TT)_{ij}\equiv q_{ij}
\end{equation}
It makes sense to change variables from $T$ to $T^T T$ because the dot products of the vectors which are the elements of the Gram matrix, are directly related to the primary invariants. A standard computation of the Jacobian gives
\begin{equation}
 \dd^6q=8r_1^3r_2^2r_3\,\dd r_1\, \dd r_2\, \dd r_3\, \dd a\,\dd b\,\dd c.
\end{equation}
Noting that $\det q=(r_1r_2r_3)^2$, we obtain the result
\begin{equation}
 \frac{\dd^9Y}{\Vol SO(3)}=\sum_{s=\pm 1}\frac18(\det q)^{-1/2}\dd^6q\label{eq:Ymeasure}
\end{equation}
where we have performed the integral over $R\in SO(3)$ on the right hand side. This takes care of ``dividing by the gauge group'' and has reduced the number of integration variables by 3.

We now include the contribution from integrating over the fourth vector. On a dense set we have
\begin{equation}
 \vec x_4=c_1\vec{x}_1+c_2\vec{x}_2+c_3\vec{x}_3=Y\vec c,  \qquad \vec c\in\mathbb R^3.
\end{equation}
Define
\begin{eqnarray}
\vec{b}=q\vec{c}, \qquad b_i=\vec{x}_i\cdot\vec{x}_4,\qquad
 r=\vec c^{\,T} q\vec{c}=\vec c\cdot \vec b
 =\vec x_4\cdot\vec x_4.
\end{eqnarray}
The full Gram matrix is given by
\begin{equation}
 G=\begin{pmatrix}q&\vec b\\ \vec b^T &r\end{pmatrix}.
\end{equation}
The linear map $\vec c\mapsto\vec x_4=Y\vec c$ has Jacobian
\begin{equation}
 \dd^3 x_4=|\det Y|\dd^3c=(\det q)^{1/2}\dd^3c.
\end{equation}
The map $\vec{c}\mapsto \vec{b}=q\vec{c}$ has Jacobian $\det q$, so
\begin{equation}
 \dd^3c=(\det q)^{-1}\dd^3b\qquad\Rightarrow\qquad \dd^3\vec x_4=(\det q)^{-1/2}\dd^3b.
\end{equation}
Multiplying by \eqref{eq:Ymeasure},
\begin{equation}
 \frac{\prod_{a=1}^4\dd^3\vec x_a}{\Vol SO(3)}=\sum_{s=\pm 1}\frac18(\det q)^{-1}\dd^6q\dd^3b.\label{eq:intermediate-measure}
\end{equation}
We want to write this in a slightly more convenient form. Towards this end, use the Schur-complement identity
\begin{equation}
 \det G=\det q\left(r-b^Tq^{-1}b\right)
\end{equation}
to find
\begin{align}
 \delta(\det G) &=\delta\!\left(\det q\,[r-b^Tq^{-1}b]\right)=\frac1{|\det q|}\,
 \delta\!\left(r-b^Tq^{-1}b\right).
\end{align}
On the positive Gram region $\det q>0$, and hence
\begin{equation}
 \int\dd r\,\delta(\det G)=\frac1{\det q}.
\end{equation}
Equation \eqref{eq:intermediate-measure} can thus be written as
\begin{equation}
 \frac{\prod_{a=1}^4\dd^3\vec x_a}{\Vol SO(3)} =\frac14\dd^{10}G\,\delta(\det G),\label{eq:grammeasure}
\end{equation}
where it is understood that on the LHS of the above equation we integrate over the $SO(3)$ gauge group and on the RHS we integrate over the coordinate $\Sigma$ introduced below. Thus on both sides of (\ref{eq:grammeasure}) we are integrating over nine coordinates. This derivation uses the dense patch $\det q=\det G_{123}>0$. The rank-two saddle encountered below lies on the boundary $\det q=0$, where these coordinates degenerate; the invariant measure formula is understood by continuation from the dense patch, while the saddle itself will be checked directly in the original matrix variables. The last step is to express the integral in terms of our chosen primary invariants. Order the independent Gram entries as
\begin{equation}
 y=(G_{11},G_{22},G_{33},G_{44},G_{12},G_{13},G_{14},G_{23},G_{24},G_{34}).
\end{equation}
Order the new invariant coordinates as $w=(p_1,\ldots,p_9,\Sigma)$. The Jacobian matrix $\mathsf J=\partial w/\partial y$ is
\begin{equation}
\mathsf J=
\begin{pmatrix}
1&0&0&0&0&0&0&0&0&0\\
0&0&1&0&0&0&0&0&0&0\\
0&-\frac14&0&-\frac14&0&0&0&0&1&0\\
0&1&0&1&0&0&0&0&0&0\\
0&1&0&-1&0&0&0&0&0&0\\
0&0&0&0&1&0&0&0&0&-1\\
0&0&0&0&0&1&0&0&0&0\\
0&0&0&0&0&0&1&0&0&0\\
0&0&0&0&0&0&0&1&0&0\\
0&0&0&0&1&0&0&0&0&0
\end{pmatrix}
\end{equation}
so that $\det\mathsf J=2$. Thus
\begin{equation}
 \dd^9p\,\dd\Sigma=2\,\dd^{10}G,
 \qquad
 \dd^{10}G=\frac12\dd^9p\,\dd\Sigma.
\end{equation}
The matrix integral we are studying now becomes
\begin{equation}
 Z(g) =\sum_{s=\pm 1}\frac{1}{16}\int\dd^9p\,\dd\Sigma\, \delta(F(\vec p,\Sigma))
 \exp\left[-\frac{\lambda}{2g}\sum_{i=1}^9p_i^2\right].\label{eq:Zdelta}
\end{equation}
If $\Sigma_r(\vec p)$ is a simple real root of $F=0$, we have the standard delta function identity
\begin{equation}
 \delta(F(\vec p,\Sigma)) =\sum_r\frac{\delta(\Sigma-\Sigma_r(\vec p))} {|\partial_\Sigma F(\vec p,\Sigma_r(\vec p))|}.
\end{equation}
Therefore
\begin{equation}
 Z(g)=\sum_{s=\pm 1}\sum_{r\in\text{physical}(\vec p)}\, \frac{1}{16}\int\dd^9p\, e^{-\lambda \vec p\cdot\vec p/(2g)} \frac1{|\partial_\Sigma F(\vec p,\Sigma_r(\vec p))|}.
\end{equation}
Above ``$\text{physical}(\vec p)$'' means we should choose only roots $\Sigma$ which lead to a Gram matrix that is obtained from Hermitian matrices. The way to test this is simply to see if the resulting Gram matrix is positive semidefinite or not. For example, this requirement rules out the roots $\Sigma=\pm\sqrt{3}$ in \eqref{examplefiber}. After complexification the absolute value is removed and one uses a holomorphic residue instead.

\section{Perturbation theory on one physical Hironaka sheet}\label{PT}

To develop a perturbation expansion, we need to work on a specific sheet. Choose the sheet satisfying $\Sigma_+(\vec 0)=1$. For definiteness we also fix one of the two orientation signs $s=\pm1$; the resulting constant normalization is absorbed into $\mathcal N_+$. Since $\partial_\Sigma F(\vec 0,1)=-4$, define the normalized sheet Jacobian
\begin{equation}
 \cJ_+(\vec p)=-\frac4{\partial_\Sigma F(\vec p,\Sigma_+(\vec p))}, \qquad\qquad \cJ_+(\vec 0)=1.\label{sheetJac}
\end{equation}
The perturbative sheet integral is
\begin{equation}
 Z_+^{\rm pert}(g) =\mathcal N_+\int_{\mathbb R^9}\dd^9p\, e^{-\lambda \vec p\cdot\vec p/(2g)}\cJ_+(\vec p).\label{PFR}
\end{equation}
In terms of the rescaled coordinate
\begin{equation}
 \vec p=\sqrt{\frac g\lambda}\,\vec z,
\end{equation}
we have
\begin{equation}
 Z_+^{\rm pert}(g) =\mathcal N_+\left(\frac{2\pi g}{\lambda}\right)^{9/2} \int_{\mathbb R^9}\frac{\dd^9z}{(2\pi)^{9/2}} \, e^{- \vec{z}\cdot\vec{z}/2}\cJ_+\left(\sqrt{\frac g\lambda}\vec z\right)\equiv \mathcal N_+\left(\frac{2\pi g}{\lambda}\right)^{9/2} 
 \left\langle \cJ_+\!\left(\sqrt{\frac g\lambda}\vec z\right) \right\rangle,
\end{equation}
where the integral over $z$ is a standard nine-dimensional Gaussian integral. Thus all nontrivial perturbation theory comes from the sheet Jacobian $\cJ_+$. Performing a power series expansion of $\cJ_+\!\left(\sqrt{\frac g\lambda}\vec z\right)$ in the variable $\vec z$ produces the perturbative expansion. Notice that odd powers of $\vec z$ vanish upon integration so that we obtain a power series in $g$. The potential itself has no interaction vertices. For a standard Gaussian integral we have
\begin{equation}
 \langle f(\sqrt{s}\,\vec z)\rangle =\left.e^{\frac{s}{2}\Delta}f(\vec p)\right|_{\vec p=0}, \qquad
 \Delta=\sum_{i=1}^9\frac{\partial^2}{\partial p_i^2}.
\end{equation}
Taking $s=g/\lambda$ in the above formula gives
\begin{equation}
 \Phi_0(g)=\frac{Z_+^{\rm pert}(g)}{\mathcal N_+\left(\frac{2\pi g}{\lambda}\right)^{9/2}} =\left. \exp\!\left(\frac{g}{2\lambda}\Delta\right)\cJ_+(\vec p) \right|_{\vec p=0}=\sum_{n=0}^\infty c_n\, g^n.
\label{eq:heat}
\end{equation}
Hence
\begin{equation}
 c_n=\frac1{2^n n!\lambda^n}\left.\Delta^n\cJ_+(\vec p)\right|_{\vec p=0}.\label{eq:cn}
\end{equation}
Thus the perturbative coefficients can be generated to arbitrary order using only differentiation. This is easily implemented in Mathematica. The first few coefficients are
\begin{equation}
c_0=1,\quad c_1=\frac{959}{128\lambda},\quad c_2=\frac{4402001}{32768\lambda^2},\quad c_3=\frac{19040130725}{4194304 \lambda ^3},\quad c_4=\frac{499392750094539}{2147483648 \lambda ^4},\quad\cdots\label{cnnums}
\end{equation}
From the first few coefficients we can see evidence that this sum is divergent.

For the first few orders, the coefficients in \eqref{eq:cn} can be obtained directly by repeated application of the nine-dimensional Laplacian.  At higher orders, however, this becomes computationally inefficient because the intermediate expressions grow very rapidly.  We therefore use an equivalent representation of the same perturbative expansion which reduces the calculation to three variables. To understand the more efficient representation, there is a useful analogy. Imagine evaluating an integral, which is constrained to a curve
\begin{equation}
F(x,y)=0.\label{curvedefined}
\end{equation}
This is much like our integral since we are integrating over the ten elements of the Gram matrix $G$ subject to the constraint $\det G=0$. To perform the integral, we could solve \eqref{curvedefined} as $y=y(x)$, in which case the measure contains $1/{\partial F\over\partial y}$. Alternatively, we can parametrize the curve directly, $(x,y)=(x(s),y(s))$, and integrate over $s$. In this case $1/{\partial F\over\partial y}$ never appears. Instead there is a Jacobian involving $dx/ds$ and $dy/ds$. 

The more efficient computation begins by solving the constraint $\det G=0$. Since the determinant of $G$ vanishes it has a non-trivial kernel spanned by its null eigenvector $\vec n$. Choose a kernel direction and work on the projective patch
\begin{equation}
\vec{n}\sim (u_1,u_2,u_3,1).\label{nullvector}
\end{equation}
A rank-three Gram matrix with this kernel can be written as
\begin{equation}
 G=\mathsf E(u)L\,\mathsf E(u)^T,\qquad \mathsf E(u)=
 \begin{pmatrix}
 1&0&0\\
 0&1&0\\
 0&0&1\\
 -u_1&-u_2&-u_3
 \end{pmatrix},
\end{equation}
where $L$ is a symmetric $3\times3$ matrix. We can relate this description to that of Section \ref{JacobQuotient}. The choice of null vector \eqref{nullvector} is equivalent to choosing 
\begin{equation}
b_a=-\sum_{b=1}^3 q_{ab}u_b\qquad r=\sum_{a,b=1}^3 u_a q_{ab}u_b\,.
\end{equation}
Thus $d^3 b=\det(q) d^3 u$ and so that using the results of Section \ref{JacobQuotient} we have
\begin{equation}
d^{10} G\,\delta(\det(G))=\frac{d^6 q d^3 b dr}{\det q}\delta (r-\vec{b}^Tq^{-1}\vec{b})=d^6 q d^3 u dr \delta (r-\vec{b}^Tq^{-1}\vec{b}).
\end{equation} 
After integrating over $r$ we are left with the trivial measure $d^6 q d^3 u$. This shows the advantage of choosing the null eigenvector as we did: the determinant factor in the constrained Gram measure is exactly cancelled by the Jacobian of the transformation $\vec b\to\vec u$. If $\vec l$ denotes the six independent entries of $L$, the nine primaries are affine-linear functions of $\vec l$,
\begin{equation}
 \vec p=A(u_a)\vec l-\vec{c}.\label{effprim}
\end{equation}
The explicit expression for the matrix $A(u_a)$ and the constants $\vec{c}$ can be obtained by comparing the above Gram matrix with the explicit formulas for the primary invariants given in \eqref{eq:udef}. Given the relation \eqref{effprim}, the quadratic form appearing in the action can be completed to a square,
\begin{equation}
\vec p\cdot\vec p=(\vec l-\vec\mu)^T K(u_a)(\vec l-\vec\mu)+\mathcal S(u_a),
\end{equation}
where
\begin{eqnarray}
K(u_a)&=&A(u_a)^T A(u_a),\qquad \qquad\vec{\mu}(u_a)\,\,=\,\,K(u_a)^{-1}A(u_a)^T\vec{c},\cr\cr
\mathcal S(u_a)&=&\vec{c}\cdot\vec{c}-\vec{c}^{\,\,T}A(u_a)K(u_a)^{-1}A(u_a)^T\vec{c}.
\end{eqnarray}
The six $L$ directions are therefore exactly Gaussian.  After carrying out these Gaussian integrations
\begin{eqnarray}
Z^{\rm pert}_+(g)&\propto&\int d^3 u \, d^6 l \exp\left[-{(A(u_a)\vec l-\vec{c})\cdot(A(u_a)\vec l-\vec{c})\over 2t}\right]\qquad (t=\frac{g}{\lambda})\cr\cr
&=&(2\pi t)^3\int d^3 u \frac{1}{\sqrt{\det K(u_a)}}\exp\left[-\frac{\mathcal S(u_a)}{2t}\right]
\end{eqnarray} 
all nontrivial perturbative information is contained in a three-dimensional integral with reduced action $\mathcal S(u_a)$ and measure proportional to $1/\sqrt{\det K(u_a)}$.  In practice it is convenient to evaluate
\begin{equation}
 {\cal D}(u_a)=\det K(u_a), \qquad \mathcal S(u_a)=\frac{{\cal N}(u_a)}{{\cal D}(u_a)},
\end{equation}
where both ${\cal N}(u_a)$ and ${\cal D}(u_a)$ are polynomials. To see why this is more efficient, we introduce the $9\times 7$ matrix
\begin{equation}
\widetilde A=(A,-\vec{c})
\end{equation}
Then
\begin{equation}
\widetilde A^{\,T}\widetilde A=\begin{pmatrix} A^T A &-A^T\vec{c}\\ -\vec{c}^T A &\vec{c}\cdot\vec{c}\end{pmatrix}
=\begin{pmatrix} K &-A^T\vec{c}\\ -\vec{c}^T A &\vec{c}\cdot\vec{c}\end{pmatrix}
\end{equation}
Using the Schur complement identity
\begin{equation}
\det (\widetilde A^{\,T}\widetilde A)=\det K\, (\vec{c}\cdot\vec{c}-\vec{c}^{\, T}AK^{-1}A^T\vec{c})
\end{equation}
Notice that the quantity in parentheses is $\mathcal S(u_a)$, so that we can identify ${\cal N}(u_a)=\det(\widetilde A^{\,T}\widetilde A)$. With this identification, we do not need to invert $K(u_a)$ to obtain $\mathcal S(u_a)$. We now want to expand about the physical vacuum
\begin{equation}
 \vec n_{\rm phys}=\frac12(1,-1,-1,1)
\end{equation}
which corresponds on this patch to
\begin{equation}
 u_0=(1,-1,-1).
\end{equation}
For the perturbative expansion only a neighborhood of $u_0$ is required. This neighborhood is a direct parametrization of the same local branch that was described above as $\Sigma=\Sigma_+(\vec{p})$; the two calculations therefore differ only in the choice of coordinates on the constrained invariant manifold. To generate the perturbative expansion, write
\begin{equation}
 u_a=(u_0)_a+\epsilon z_a, \qquad \epsilon^2=t=\frac{g}{\lambda}.
\end{equation}
We integrate over the three coordinates packaged in $z$. To generate the perturbative expansion of the integrand
\begin{equation}
\frac{1}{\sqrt{{\cal D}(u_a)}}\exp\left(-\frac{\mathcal S(u_a)}{2t}\right)=\frac{1}{\sqrt{{\cal D}(u_a)}}\exp\left(-\frac{{\cal N}(u_a)}{2t{\cal D}(u_a)}\right)
\end{equation}
Since we expand about a minimum we can define a new function ${\cal N}((u_0)_a+\epsilon z_a)=\epsilon^2 M(\epsilon,z_a)$. In terms of
\begin{equation}
\frac{M(0,z)}{2{\cal D}(0)}=\frac12 \sum_{a,b=1}^3 z_a H_{ab} z_b,\qquad\qquad {\cal D}_0={\cal D}(u_0),\qquad\qquad {\cal D}_\epsilon={\cal D}((u_0)_a+\epsilon z_a),
\end{equation}
we can factor out the leading Gaussian with covariance $C_{ab}=H_{ab}^{-1}$, where $H$ is the Hessian of $S/2$ at $u_0$ to obtain
\begin{equation}
\frac{1}{\sqrt{{\cal D}(u_a)}}\exp\left(-\frac{\mathcal S(u_a)}{2t}\right)=\frac{e^{-\frac12 \sum_{a,b=1}^3 z_a H_{ab} z_b}}{\sqrt{{\cal D}_0}}R(\epsilon,z_a),
\end{equation}
with
\begin{equation}
 R(\epsilon,z_a) =\sqrt{{\cal D}_0\over {\cal D}_\epsilon}\exp\left[-\frac{M(\epsilon,z)}{2{\cal D}_\epsilon}+\frac{M(0,z)}{2{\cal D}_0}\right]= 1+\sum_{m\geq1}r_m(z)\epsilon^m.
\end{equation}
Rather than expanding the exponential directly, we exploit the fact that $R$ obeys a first-order differential equation in $\epsilon$,
\begin{equation}
 P(\epsilon,z_a)\,\partial_\epsilon R(\epsilon,z_a) + Q(\epsilon,z_a)\,R(\epsilon,z_a)=0,\label{DeqnP}
\end{equation}
where $P$ and $Q$ are finite polynomials determined algebraically by ${\cal N}$ and ${\cal D}$
\begin{equation}
P(\epsilon,z_a)=2{\cal D}_\epsilon^2\qquad Q(\epsilon,z_a)=({\cal D}_\epsilon-M(\epsilon,z_a)){\cal D}'_\epsilon+M'(\epsilon,z_a){\cal D}_\epsilon
\end{equation} 
The differential equation \eqref{DeqnP} is easily derived by differentiating
\begin{equation}
\log R(\epsilon,z)=-\frac12\log {\cal D}_\epsilon - \frac{M(\epsilon,z)}{2{\cal D}_\epsilon}+\cdots
\end{equation} 
where $\cdots$ stand for terms independent of $\epsilon$. Substituting the power series for $R$ therefore gives a finite recurrence which determines $r_m$ from a fixed number of preceding coefficients.  This avoids the rapidly growing symbolic expressions produced by repeated application of $\Delta$.

Finally, only the even powers contribute after Gaussian integration.  The perturbative coefficients are obtained from
\begin{equation}
 d_n\equiv \lambda^n c_n = \big\langle r_{2n}(z)\big\rangle_H,
\end{equation}
where the brackets denote the Gaussian average with covariance $H^{-1}$.  These averages are evaluated algebraically using Wick contractions.  As a check, the reduced computation reproduces
\begin{equation}
 d_1=\frac{959}{128}, \qquad d_2=\frac{4402001}{32768}, \qquad
 d_3=\frac{19040130725}{4194304}, \qquad d_4=\frac{499392750094539}{2147483648},
\end{equation}
in agreement with \eqref{cnnums}.  The advantage of this formulation is that arbitrarily high perturbative orders can be generated without solving explicitly for the algebraic root $\Sigma_+(\vec p)$ and without repeatedly applying the nine-dimensional Laplacian. We will use the first 50 orders of perturbation theory in Section \ref{sec:hessian-one-loop} to identify the Borel--Leroy singularity. Although the coefficients $d_n$ for $n=1,2,\cdots,50$ are too lengthy to quote here, they are incuded as a text file with the arXiv submission of this paper.

\section{Exact angular-integral representation of Borel--Leroy transform}\label{BorelLeroy}

In this section we derive an exact angular-integral representation of the Borel-Leroy transform of the perturbation theory series. As explained in Appendix \ref{BorelSing}, the location of the singularities of the Borel-Leroy transform is identical to the location of the singularities of the Borel transform. This representation therefore allows us to search for singularities in the Borel transform itself. To obtain the angular-integral representation, notice that the integral in \eqref{PFR}, after expanding ${\cal J}_+$ in a power series, is a sum of moments of a Gaussian integral in 9 dimensions. By changing to a radius and angles the radial piece of the resulting integral takes the form of a Laplace transform of a series in the coupling. Consequently, after doing the angular integrations, the result is naturally interpreted as the Borel-Leroy transform of the perturbation series. We make this intuition precise below. 

The quotient has nine continuous coordinates, so after changing to a radius and angles the natural radial power is nine-dimensional. Introduce ordinary polar coordinates in the primary invariant space
\begin{equation}
\vec p=r\widehat n, \qquad\qquad \widehat n\in S^8, \qquad\qquad \dd^9p=r^8\dd r\,\dd\Omega_8.
\end{equation}
The potential is $V=\frac\lambda2r^2$. Define the Borel variable
\begin{equation}
 \zeta=V=\frac\lambda2r^2,\qquad r=\sqrt{\frac{2\zeta}{\lambda}}, \qquad 
 r^8\dd r=\frac12\left(\frac2\lambda\right)^{9/2} \zeta^{7/2}\dd\zeta.
\end{equation}
\emph{The Borel variable is literally the potential value $\zeta=V(\vec p)$.} After dividing by the Gaussian normalization one obtains
\begin{equation}
 \Phi_0(g)= \frac1{\Gamma(9/2)g^{9/2}}\int_0^\infty\dd\zeta\, e^{-\zeta/g}\zeta^{7/2} B_{9/2}[\Phi_0](\zeta),
\end{equation}
with
\begin{equation}
B_{9/2}[\Phi_0](\zeta) =\frac1{\Omega_8}\int_{S^8}\dd\Omega_8(\widehat n)\,\cJ_+\!\left(\sqrt{\frac{2\zeta}{\lambda}}\,\widehat n\right).\label{eq:BorelAngular}
\end{equation}
A Borel singularity can arise when the sphere of radius $\sqrt{2\zeta/\lambda}$ encounters a singularity of the sheet projection, i.e. a zero of $\partial_\Sigma F$ compatible with the full saddle conditions.
To obtain a more explicit formula for $B_{9/2}[\Phi_0](\zeta)$ we need to perform the angular integrations appearing in \eqref{eq:BorelAngular}. This is accomplished with the following rules for angular integration
\begin{eqnarray}
\frac1{\Omega_8}\int_{S^8}\dd\Omega_8(\widehat n)\,\, \hat{n}_{i_1}\hat{n}_{i_2}\cdots\hat{n}_{i_{2k+1}}&=&0,\cr\cr
\frac1{\Omega_8}\int_{S^8}\dd\Omega_8(\widehat n)\,\, \hat{n}_{i}\hat{n}_{j}&=&\frac{\delta_{ij}}{9},\cr\cr
\frac1{\Omega_8}\int_{S^8}\dd\Omega_8(\widehat n)\,\, \hat{n}_{i_1}\hat{n}_{i_2}\cdots\hat{n}_{i_{2m}}&=&\frac{\sum_{\rm pairs}\delta_{\rm pair\, 1}\cdots\delta_{\rm pair\, m}}{9\cdot 11\cdot 13\cdots (9+2m-2)}.
\end{eqnarray}
Using these rules we easily find
\begin{equation}
B_{9/2}[\Phi_0](\zeta) =\sum_{n=0}^\infty c_n \,\frac{\Gamma(9/2)}{\Gamma(n+9/2)}\,\zeta^n
\end{equation}
which is the Borel-Leroy transform of the perturbation series. This proves that \eqref{eq:BorelAngular} does indeed provide an exact angular-integral representation of the Borel-Leroy transform. We use this representation to search for singularities of the Borel transform in the next section.

\section{Ramification and a candidate leading saddle}\label{Ram}

The 8-sheeted Hironaka cover over the primary base is ramified: different sheets can meet over the same primary point. The equation $F(\vec p,\Sigma)=0$ describes the four-sheeted $\Sigma$-subcover. Concretely, $F(\vec p,\Sigma)$ is given by
\begin{equation}
F(\vec{p},\Sigma)=(\Sigma-\Sigma_1(\vec p))(\Sigma-\Sigma_2(\vec p))(\Sigma-\Sigma_3(\vec p))(\Sigma-\Sigma_4(\vec p)),
\end{equation}
where we used the fact that \eqref{eq:G} implies the coefficient of $\Sigma^4$ in $F(\vec p,\Sigma)$ is 1. At a smooth point of the hypersurface $F(\vec p,\Sigma)=0$, the projection to primary space is ramified when
\begin{equation}
F(\vec p,\Sigma)=0,\qquad \frac{\partial F}{\partial\Sigma}(\vec p,\Sigma)=0.\label{sing}
\end{equation}
The rank-two saddle encountered below is more singular: all first derivatives of $F$ vanish there, as explained in Section~\ref{sec:ranktwosaddle}. It is clear from \eqref{sheetJac} that the sheet Jacobian $\cJ_+(\cdot)$ is singular at these ramification points \eqref{sing}, and hence from \eqref{eq:BorelAngular} we may get a singularity of the Borel-Leroy transformation. In this section our goal is to test this intuitive observation linking sheet collisions and singularities of the Borel--Leroy transform of the perturbative series, and to see it working in practice. 

\subsection{Why a rank-two sheet collision can become a saddle}
\label{sec:ranktwosaddle}

The perturbative expansion on a given Hironaka sheet becomes singular when the projection from the complex cover to primary space ceases to be locally one-to-one. The saddle that we find below occurs at a particularly interesting point of the cover: the corresponding Gram matrix has rank two.  The geometry near such a point is more singular than the geometry of a generic simple ramification.  In this subsection we explain this geometry and why it naturally produces a saddle of the matrix integral.

In the original matrix variables the integral has the form
\begin{equation}
 Z(g)=\int\prod_{a=1}^4 dX^a\, \exp\left[-\frac{V(X)}{g}\right],\qquad V(X)=\frac{\lambda}{2}\sum_{i=1}^9p_i(X)^2 . \label{eq:potential-original}
\end{equation}
The natural ``Euclidean action'' is
\begin{equation}
 S_E(X)=\frac{1}{g}V(X).
\end{equation}
Since the overall factor $1/g$ does not affect the saddle-point equations, we can study critical points of $V$ directly. In terms of the primary invariants, recall that
\begin{equation}
 V(\vec p)=\frac{\lambda}{2}\sum_{i=1}^9p_i^2. \label{eq:potential-primary}
\end{equation}

We now ask what is special about a saddle that corresponds to a rank-two Gram matrix.  For any matrix $G$,
\begin{equation}
 \delta\det G\,\,=\,\, {\rm Tr}\left({\rm adj}(G)\,\delta G\right), \label{eq:detvariation}
\end{equation}
where ${\rm adj}(G)$ is the adjugate matrix.  If $G$ is a generic rank-three $4\times4$ Gram matrix, ${\rm adj}(G)$ has rank one and is nonzero. Consequently the hypersurface $\det G=0$ is smooth at a generic rank-three point. The situation changes qualitatively when ${\rm rank}\,G=2$. Every $3\times3$ minor of $G$ then vanishes and hence
\begin{equation}
 {\rm adj}(G)=0\qquad\Rightarrow\qquad \delta\det G=0
\end{equation}
for every infinitesimal variation of $G$.  Since the variables $(p_1,\ldots,p_9,\Sigma)$ are linear combinations of the Gram coordinates, it follows that at a rank-two point
\begin{equation}
 F=0,  \qquad \frac{\partial F}{\partial\Sigma}=0, \qquad \frac{\partial F}{\partial p_i}=0,
 \qquad i=1,\ldots,9. \label{eq:ranktwo-singular}
\end{equation}
Thus a rank-two point is not merely a generic point at which two roots of $F$ happen to coincide.  It is a singular point of the hypersurface $F=0$ itself\footnote{A point on this hypersurface is smooth provided the gradient of F is not zero: $\nabla F=\left({\partial F\over\partial p_1},\cdots,{\partial F\over\partial p_9},{\partial F\over\partial \Sigma}\right)\ne 0$. At a generic collision of two $\Sigma$-sheets we have $F=0$ and $\partial_\Sigma F=0$. This does not imply that $\partial_{p_i}F=0$ for all $i$. As long as at least one $\partial_{p_i}F\ne 0$, we have $\nabla F\ne 0$ and the hypersurface $F=0$ is perfectly smooth. What has become singular is only the projection $(\vec{p},\Sigma)\mapsto \vec{p}$.}.

Another way to see that the cover is degenerating more strongly at such a point, is that rank two implies that every $3\times3$ Gram determinant vanishes.  In particular,
\begin{equation}
 D(\vec p,\Sigma)=\det G_{123}=0,
\end{equation}
and hence $\Omega^2=D=0$. All four cubic orientation invariants vanish because the four vectors lie in a common plane.  Thus the orientation sheets meet at the same point.  The rank-two saddle is a higher sheet-collision point of the full Hironaka cover, rather than a generic simple two-sheet ramification.

It is instructive to consider the local geometry in the original vector variables.  At a rank-two configuration the four vectors span a two-dimensional plane.  Let $\hat n$ be a unit vector normal to this plane, i.e. $\vec x_a\cdot\hat n=0$ for all $a$.  Consider a deformation normal to the plane,
\begin{equation}
 \vec x_a(\epsilon) = \vec x_a+\epsilon w_a\hat n, \qquad a=1,\ldots,4. \label{eq:normaldeformation}
\end{equation}
The corresponding Gram matrix is
\begin{equation}
 G_{ab}(\epsilon) = \vec x_a(\epsilon)\cdot\vec x_b(\epsilon)
 = \vec x_a\cdot\vec x_b+\epsilon w_b\,\vec x_a\cdot\hat n+\epsilon w_a\,\hat n\cdot\vec x_b+\epsilon^2w_aw_b= G_{ab}(0)+\epsilon^2w_aw_b. \label{eq:gramquadratic}
\end{equation}
Note that there is \emph{no term linear in $\epsilon$}. A deformation which moves the configuration out of the rank-two plane is invisible to the Gram invariants to first order.

Because each primary $p_i$ is an affine-linear function of the Gram entries, equation \eqref{eq:gramquadratic} implies
\begin{equation}
 p_i(\epsilon) = p_i^\star+\epsilon^2 q_i(w), \label{eq:primaryquadratic}
\end{equation}
where $q_i(w)$ is quadratic in the four numbers $w_a$.  Consequently the potential behaves as
\begin{align}
 V(\epsilon) &= \frac{\lambda}{2} \sum_i \left(p_i^\star+\epsilon^2q_i(w)\right)^2
= V_\star +\lambda\epsilon^2 \sum_i p_i^\star q_i(w) +O(\epsilon^4). \label{eq:potentialnormal}
\end{align}
In particular,
\begin{equation}
 \frac{dV}{d\epsilon}\bigg|_{\epsilon=0}=0. \label{eq:automaticstationarity}
\end{equation}
Thus stationarity in the directions that move the vectors out of the rank-two plane is automatic.  This is the rank-two analogue of the fact that a ramified coordinate can enter the primaries only quadratically. For the rank-two saddle there is more than one such physical direction. We can count these directions explicitly.  Before quotienting by rotations there are four independent numbers $w_1,w_2,w_3,w_4$ in \eqref{eq:normaldeformation}.  Two combinations, however, are pure gauge.  Indeed, an infinitesimal rotation acts as
\begin{equation}
 \delta\vec x_a=\vec\omega\times\vec x_a.
\end{equation}
If $\vec\omega$ lies in the plane spanned by the $\vec x_a$, then $\vec\omega\times\vec x_a$ is normal to the plane.  There are two independent choices of $\vec\omega$ inside the plane so two of the four normal deformations in \eqref{eq:normaldeformation} are simply infinitesimal $SO(3)$ gauge rotations.  After quotienting by rotations, therefore, $4-2=2$ physical transverse directions remain.

This counting can also be understood more directly: the generic quotient has dimension nine.  A rank-two configuration can be represented by four vectors in a plane.  Using $SO(3)$, place the two-dimensional span of the four vectors in a fixed reference plane. The four planar vectors then have eight real parameters, and the residual $SO(2)$ rotation removes one, so such configurations have
\begin{equation}
 4\times2-\dim SO(2)=8-1=7
\end{equation}
continuous invariant parameters.  The rank-two locus therefore has dimension seven inside the nine-dimensional quotient, leaving two transverse directions.  These are the two physical directions whose effect on the primary invariants starts at quadratic order.

We can now state the saddle mechanism.  Near a generic rank-two point the physical variations split into two classes:
 7 directions (tangent to the rank-two locus) and 2 directions (transverse and ramified). The first derivatives of the potential in the two transverse directions vanish automatically, as shown in \eqref{eq:automaticstationarity}.  To obtain a genuine critical point, we must therefore require that the potential also be stationary in the seven directions tangent to the rank-two locus
\begin{equation}
 d\!\left(V\big|_{{\rm rank}\,G=2}\right)=0. \label{eq:ranktwostationarity}
\end{equation}
This is the nontrivial part of the saddle-point condition.

To summarize this discussion, at a generic rank-three point the Gram invariants provide good local coordinates on the quotient.  At rank two, two physical directions become invisible to the invariants at first order.  The Gram hypersurface becomes singular and the projection of the finite Hironaka cover to primary space loses rank in two physical directions, several sheets meet over the same primary point and the potential is automatically stationary in the two transverse directions.  If the restriction of the potential to the seven-dimensional rank-two locus is also stationary, the resulting point is a candidate saddle of the original matrix integral.

The invariant-space argument explains why a rank-two sheet collision \emph{might} produce a saddle. What is missing is that it does not prove a particular solution is stationary under every variation of the original matrix variables.  We will derive the full matrix saddle equation in the next subsection and use it as a direct check of the candidate saddle.

\subsection{Full matrix saddle equation}

Let $\mathbf X$ be the $3\times4$ matrix whose columns are the vectors $\vec x_a$.  Since $G=\mathbf X^T\mathbf X$, write
\begin{equation}
 \delta V=\lambda\Tr(B\,\delta G)
\end{equation}
for a symmetric $4\times4$ matrix $B$.  Then
\begin{equation}
\delta G=\mathbf X^T\delta\mathbf X+(\delta\mathbf X)^T\mathbf X,\qquad\Rightarrow\qquad \delta V=2\lambda\Tr[(\mathbf X B)^T\delta\mathbf X].
\end{equation}
Hence the condition for a stationary point of the potential is
\begin{equation}
\mathbf X B=0.\label{eq:XB}
\end{equation}
On the real positive-semidefinite slice this implies $GB=0$ and is equivalent to it whenever $G=\mathbf X^T\mathbf X$. For the potential \eqref{eq:Vsimple}, the entries of $B$ are easy to obtain by the chain rule.  For example
\begin{equation}
 B_{11}=p_1, \qquad B_{13}=\frac12p_7, \qquad B_{12}=\frac12p_6,
\end{equation}
with the remaining entries obtained similarly from \eqref{eq:udef}.  This is the direct test that a candidate discriminant critical point is a genuine saddle of the original matrix integral.

\subsection{Symmetry reduction}\label{SymRed}

We now want to explicitly demonstrate a saddle at a ramification. In this section we make a symmetry-motivated ansatz that reduces the dimensionality of the saddle search. The relevant symmetry is the $\mathbb Z_2$ permutation $1\leftrightarrow 3$, $2\leftrightarrow 4$ acting on the four vectors as $(\vec x_1,\vec x_2,\vec x_3, \vec x_4)\to (\vec x_3,\vec x_4, \vec x_1,\vec x_2)$. Under this permutation the primaries transform as
\begin{equation}
p_1\leftrightarrow p_2,\quad p_3\to p_3,\quad p_4\to p_4, \quad p_5\to -p_5,\quad p_6\to -p_6,\quad p_7\to p_7,\quad p_8\leftrightarrow p_9.
\end{equation}
A simple $\mathbb Z_2$ invariant ansatz is given by
\begin{equation}
 p_1=p_2=x, \quad p_3=w, \quad p_4=v, \quad p_5=p_6=0, \quad p_7=z, \quad p_8=p_9=y.\label{eq:ansatz}
\end{equation}
On this subspace
\begin{align}
 F=\frac1{16}P_+P_-,\qquad{\rm where}\qquad P_+&=-4(\Sigma+y)^2+(1+x+z)(12+3v+4w),\\
P_-&=-4(\Sigma-y)^2+(1+x-z)(4+v-4w).
\end{align}
The potential evaluated on this ansatz is
\begin{equation}
 V =\lambda(x^2+\frac12w^2+\frac12v^2+\frac12z^2+y^2).\label{eq:ansatzV}
\end{equation}
A rank-two collision on this factorized locus obeys $P_+=P_-=0$. To impose stationarity, one approach is to introduce the Lagrange multipliers $\alpha,\beta$ and solve
\begin{equation}
 \nabla\left(\frac{V}{\lambda}\right) =\alpha\nabla P_++\beta\nabla P_-, \qquad P_+=P_-=0,\label{LMC}
\end{equation}
where the gradient acts on $(x,w,v,z,y,\Sigma)$. In this approach we do not require $\nabla\frac{V}{\lambda}=0$ in primary space. Alternatively we can solve the two constraints explicitly and extremize the resulting restricted potential. The resulting candidate is subsequently checked against the full matrix equation \eqref{eq:XB}. We will use this second approach. The lowest positive-action solution we find in this symmetry sector is (see Appendix \ref{statpoint} for details)
\begin{align}
x&=-0.03840628, & w&=-0.11738894,\\
v&=-0.00929474, & z&=-0.22483345,\\
y&=+0.15268396, & \Sigma_\star&=1.30287686.
\end{align}
Substitution into \eqref{eq:ansatzV} gives
\begin{equation}
 V_\star=0.0569957516133\,\lambda.\label{eq:Astar}
\end{equation}
The corresponding Gram matrix has eigenvalues
\begin{equation}
 {\rm eig}G_\star \simeq \{0,0,2.30149242,3.61240028\},
\end{equation}
so the point is genuinely rank two. A direct evaluation of $GB$ gives zero to numerical precision (norm below $10^{-13}$).  Thus this point is not merely a stationary point of a restricted discriminant ansatz: it passes the full matrix saddle test \eqref{eq:XB}.

It is useful to compare the remarkably small value $V_\star\simeq 0.057\,\lambda$ with the scales obtained from two simple one-dimensional deformations of the physical minimum. First consider
\begin{equation}
\vec x_1=q\vec e_1,\qquad
\vec x_2=\vec e_1+\vec e_3,\qquad
\vec x_3=q\vec e_2,\qquad
\vec x_4=\vec e_2+\vec e_3.
\end{equation}
Taking the path to run from $q=1$ to $q=-1$ we transition between the two Gram roots $\Sigma=\pm 1$, while the chosen cubic orientation invariant $\Omega=T_{123}$ has the same value at the two endpoints. Along this path only $p_1=p_2=q^2-1$ are nonzero, so the potential reduces to
\begin{equation}
V(q)=\lambda(q^2-1)^2.
\end{equation}
The barrier along this particular path is therefore $V_G=\lambda$. As a second comparison, uniformly scale the physical minimum, $X_a\longrightarrow zX_a$. Taking the path to run from $z=1$ to $z=-1$ we transition between the two orientations of the cubic invariants, returning to the same Gram matrix, and hence the same value of $\Sigma$, at the two endpoints. Then
\begin{equation}
p_1=p_2=z^2-1,\qquad p_4=4(z^2-1),
\end{equation}
with all other primaries vanishing, and hence
\begin{equation}
V(z)=9\lambda(z^2-1)^2.
\end{equation}
The corresponding barrier along this radial path is\footnote{It is important to stress that $V_G$ and $V_O$ are barriers only along the particular one-dimensional paths chosen above. We do not claim that they are saddle actions of the full multidimensional problem. Their role is simply to provide a scale against which the much smaller value of $V_\star$ can be appreciated.} $V_O=9\lambda$. Both of these scales are much larger than $V_\star\simeq0.057\,\lambda$. This comparison illustrates why the ramification analysis is useful. This low-action saddle is not visible from the most obvious one-dimensional deformations of the original matrix variables; instead, it is revealed by the finite-cover geometry of the invariant description.

\subsection{Hessian, one-loop factor and global Picard--Lefschetz data}
\label{sec:hessian-one-loop}

In the previous subsection we identified a nontrivial rank-two saddle. The perturbative minimum has $V_0=0$. Define the potential difference
\begin{equation}
A_\star\equiv V_\star-V_0=V_\star=0.0569957516133\,\lambda .\label{eq:Astar-64}
\end{equation}
If this saddle controls the leading nonperturbative correction to the perturbative expansion about the physical minimum, then the saddle action $A_\star$ predicts the location of the associated Borel singularity. In this subsection we determine its local one-loop normalization and global Picard--Lefschetz connection coefficient, which fix the strength of the singularity, and we independently confirm the predicted nonperturbative scale from the large-order growth of perturbation theory.

A saddle contributes both the exponential factor $e^{-A_\star/g}$ and a Gaussian fluctuation determinant coming from quadratic fluctuations around the saddle. Comparing this fluctuation factor with the corresponding factor around the perturbative minimum gives the one-loop normalization of the nonperturbative sector. We compute this local factor and then consider the global question of how the perturbative thimble is connected to the thimble of the rank-two saddle.

Begin with the quadratic fluctuations. Let $X_0$ denote the perturbative minimum and $X_\star$ the nontrivial saddle. Expanding the potential around either critical point gives
\begin{equation}
V(X_\alpha+\delta X)=V(X_\alpha)+\frac12\,\delta X^{T}\mathcal H_\alpha\,\delta X+\cdots,\qquad
\alpha=0,\star ,\label{eq:hessian-expansion-64}
\end{equation}
where $\mathcal H_\alpha$ is the Hessian of the potential with respect to the original twelve real vector components. A Gaussian integral over nonzero fluctuations produces a factor proportional to
\begin{equation}
\frac{1}{\sqrt{\det{}'\mathcal H_\alpha}},
\end{equation}
where the prime indicates that exact zero modes have been omitted. The ratio of Gaussian fluctuation factors contains
\begin{equation}
\sqrt{\frac{\det{}'\mathcal H_0}{|\det{}'\mathcal H_\star|}}.\label{eq:hessian-ratio-64}
\end{equation}
The absolute value is required at the saddle because, as we will explicitly find below, the Hessian has one negative eigenvalue. This is an index-one saddle\footnote{An index-1 saddle means that, after removing any exact zero modes from symmetries, the Hessian of the potential has exactly one negative eigenvalue. The Morse index of a nondegenerate critical point is the number of negative eigenvalues of the Hessian there. Thus our saddle has Morse index 1.}: there is one unstable direction, while all remaining physical directions are locally stable.

The original vector space has twelve real coordinates. Three Hessian zero modes at a generic configuration arise from infinitesimal simultaneous $SO(3)$ rotations of all four vectors. Removing these three rotational zero modes leaves nine nonzero eigenvalues. At the physical minimum we find
\begin{equation}
\det{}'\mathcal H_0=229376\,\lambda^9 .\label{eq:hessian-minimum-64}
\end{equation}
At the rank-two saddle, the nine nonzero eigenvalues are approximately
\begin{align}
\frac{1}{\lambda}\,{\rm eig}'\mathcal H_\star\simeq\{&-0.378930,\, 0.305515,\, 2.383371,\, 2.622783,\, 4.130717,\nonumber\\
&5.342780,\, 7.596911,\, 15.986854,\,18.386925\}.\label{eq:hessian-eigs-64}
\end{align}
so that
\begin{equation}
|\det{}'\mathcal H_\star|\simeq 35665.404\,\lambda^9 . \label{eq:hessian-saddle-64}
\end{equation}
This confirms that the saddle has Morse index one on the quotient by rotations.

Now consider the rotational zero modes and the metric on $SO(3)$ orbits. The three zero modes must be treated separately. They correspond to moving along the $SO(3)$ orbit
\begin{equation}
\vec x_a\longrightarrow R\vec x_a,\qquad R\in SO(3),
\end{equation}
which leaves all Gram invariants, and hence the potential, unchanged. Recall that $\mathbf X$ is the $3\times4$ matrix whose columns are the four vectors. For an infinitesimal rotation $\delta\mathbf X = \omega\times\mathbf X$ and the induced metric on the three rotational collective coordinates is (see Appendix \ref{OrbMet})
\begin{equation}
M_{\rm orb}= \bigl({\rm Tr}\mathbf X\mathbf X^T\bigr)I_3-\mathbf X\mathbf X^T .\label{eq:orbit-metric-64}
\end{equation}
The factor $\sqrt{\det M_{\rm orb}}$ measures the local volume density along the rotational orbit. Since the orbit geometry at the perturbative minimum and at the saddle is not identical, the ratio of these collective-coordinate measures contributes to the one-loop prefactor.

For the two configurations we find
\begin{equation}
\det M_{{\rm orb},0}=56,\qquad\det M_{{\rm orb},\star}\simeq 49.16758284 .\label{eq:orbit-determinants-64}
\end{equation}
Combining the nonzero Hessian fluctuations with the rotational collective-coordinate factor gives the normalized one-loop ratio
\begin{equation}
R_{\rm 1loop}=\sqrt{\frac{\det M_{{\rm orb},\star}}{\det M_{{\rm orb},0}}}\sqrt{\frac{\det{}'\mathcal H_0}{|\det{}'\mathcal H_\star|}}.\label{eq:R1loop-definition-64}
\end{equation}
Numerically,
\begin{equation}
R_{\rm 1loop}\simeq 2.37627024 .\label{eq:R1loop-value-64}
\end{equation}
Using this local analysis around the saddle we therefore predict not only the exponential scale $A_\star$, but also the relative Gaussian normalization of the corresponding nonperturbative sector.

Next we consider the Picard--Lefschetz information, which controls the Borel singularity. There are two integers which play different roles. Let $\mathcal J_0$ and $\mathcal J_\star$ denote the downward-flow thimbles attached to the perturbative minimum and the rank-two saddle. Across the Stokes ray on the positive real $g$ axis the perturbative thimble may jump according to
\begin{equation}
\mathcal J_0^+=\mathcal J_0^-+m_{0\star}\,\mathcal J_\star .\label{eq:PL-jump-64}
\end{equation}
The integer $m_{0\star}$ is the \emph{Stokes connection coefficient}: up to orientation it counts complete gradient-flow trajectories connecting the two critical points. This integer enters the Stokes discontinuity of perturbation theory about $X_0$ and controls the coefficient of the Borel singularity. This should be distinguished from the coefficient of $\mathcal J_\star$ in the decomposition of the full physical integration contour, which is denoted $\nu_\star$ below.

Although the exact kernel representation of the Gram integral is a useful reduced description of the saddle geometry, at a rank-two point the null vector is not unique, so it cannot be used to count full Picard--Lefschetz trajectories. We will use the kernel reduction below to reproduce the saddle action, but we determine $m_{0\star}$ directly from the gradient flow in the original matrix variables. Let $\vec y$ denote the ten independent entries of the symmetric Gram matrix. Since the primaries are affine-linear functions of the Gram entries, we have
\begin{equation}
\vec p=L\,\vec y-\vec c .
\end{equation}
Choose a unit vector $\vec n\in S^3$ and impose that it is a null vector of the Gram matrix,
\begin{equation}
G\vec n=0.
\end{equation}
For fixed $\vec n$ these are four linear constraints on $\vec y$. Define the reduced function
\begin{equation}
\mathcal S(\vec n)\equiv\min_{\substack{G=G^T\\ G\vec n=0}} \sum_{i=1}^9p_i(G)^2 .\label{eq:kernel-effective-64}
\end{equation}
The minimization is a constrained quadratic problem, so for fixed $\vec n$ the remaining Gram variables are Gaussian and have a unique stationary value. The potential evaluated at this stationary value is
\begin{equation}
W(\vec n)=\frac{\lambda}{2}\,\mathcal S(\vec n).\label{eq:kernel-W-64}
\end{equation}
The generic space of kernel directions is $S^3/\mathbb Z_2\simeq{\mathbb R}P^3$ because $\vec n$ and $-\vec n$ define the same kernel line. For a rank-three Gram matrix the kernel is one-dimensional and the kernel line is locally unique. At the rank-two saddle, however, $\dim\ker G_\star=2$, so the fiber of the map $(G,[\vec n])\mapsto G$ is $\mathbb P(\ker G_\star)\simeq{\mathbb R}P^1$. Consequently different reduced kernel representatives can describe the same full matrix saddle, and reduced Morse trajectories need not be distinct Picard--Lefschetz trajectories in the original configuration space.

Solving the reduced stationary-point equations gives, up to the identification $\vec n\sim-\vec n$,
\begin{equation}
\vec n_\star\simeq
(-0.14058879,\,-0.10158599,\,-0.79825769,\,0.57680130).\label{eq:kernel-nstar-64}
\end{equation}
At this point
\begin{equation}
\mathcal S(\vec n_\star)=0.11399150322664,
\end{equation}
and hence
\begin{equation}
W(\vec n_\star)=\frac{\lambda}{2}\mathcal S(\vec n_\star)=0.05699575161332\,\lambda=A_\star .
\end{equation}
Thus the kernel reduction reproduces the saddle action found above using a symmetry-reduced calculation. The Hessian of $\mathcal S$ in the three tangent directions to $S^3$ at $\vec n_\star$ has eigenvalues
\begin{equation}
{\rm eig}\,{\rm Hess}_{S^3}\mathcal S(\vec n_\star)\simeq\{-0.477794,\ 12.823395,\ 29.514682\}.
\end{equation}
There is again exactly one negative eigenvalue. The unstable manifold of the reduced saddle is therefore one-dimensional and, after removing the saddle point itself, has two branches. One branch ends at
\begin{equation}
\vec n_{\rm phys}=\frac12(1,-1,-1,1),\qquad\mathcal S(\vec n_{\rm phys})=0,\qquad \Sigma=1,
\end{equation}
which is the physical perturbative vacuum. Numerically integrating the decreasing-$\mathcal S$ gradient flow from 
$\vec n_\star$
\begin{equation}
\dot n=-(I-nn^T)\nabla\mathcal S,
\end{equation} 
with initial conditions displaced infinitesimally along the two signs of the unique negative Hessian eigenvector
\begin{equation}
n_\pm(0)=\frac{n_\star\pm\epsilon v_-}{|n_\star\pm\epsilon v_-|},
\end{equation} 
we find that the two branches have different endpoints. One branch converges to
\begin{equation}
\vec n_{\rm phys}=\frac12(1,-1,-1,1),\qquad\mathcal S(\vec n_{\rm phys})=0,\qquad \Sigma=1,
\end{equation}
which is the physical perturbative vacuum. The other branch converges to
\begin{equation}
\vec n_{\rm alg}=\left(\sqrt{\frac38},-\frac1{\sqrt8},\sqrt{\frac38},-\frac1{\sqrt8}\right),\qquad
\mathcal S(\vec n_{\rm alg})=0,\qquad \Sigma=\sqrt3 .
\end{equation}
The latter is the algebraic zero-action solution which does not belong to the positive-semidefinite Hermitian slice. Thus the reduced Morse flow has the structure
\begin{equation}
\Sigma=1\ {\rm physical\ vacuum}\quad\longleftarrow\quad X_\star \quad\longrightarrow\quad
\Sigma=\sqrt3\ {\rm algebraic\ vacuum}.
\end{equation}
This flow is a statement about the auxiliary reduced function $\mathcal S(\vec n)$. To determine the Stokes incidence directly, return to the full matrix saddle equation. Equation~\eqref{eq:XB} implies
\begin{equation}
\frac{d\mathbf X}{d\tau}=-\nabla_{\mathbf X}V=-2\lambda\,\mathbf X B ,
\label{eq:full-gradient-flow}
\end{equation}
so that the Gram matrix obeys the closed equation
\begin{equation}
\frac{dG}{d\tau}=-2\lambda\,(BG+GB).
\label{eq:gram-gradient-flow}
\end{equation}
The full Hessian at $X_\star$ has one negative physical eigenvalue, so the unstable manifold of the full saddle is one-dimensional and has two branches. In an $SO(3)$ frame in which the rank-two saddle lies in the first two spatial directions, a normalized negative mode may be chosen as
\begin{equation}
\delta\mathbf X_-\simeq
\begin{pmatrix}
0&0&0&0\\
0&0&0&0\\
-0.630890993&0.319337682&-0.630890993&0.319337682
\end{pmatrix},
\end{equation}
\begin{equation}
\mathcal H_\star\,\delta\mathbf X_-\simeq-0.378929947\,\lambda\,\delta\mathbf X_- .
\label{eq:negative-full-mode}
\end{equation}
Consider the two initial conditions $\mathbf X_\pm(0)=\mathbf X_\star\pm\epsilon\,\delta\mathbf X_-$. Since the saddle is planar and $\delta\mathbf X_-$ is normal to that plane, $\mathbf X_\star^T\delta\mathbf X_-=0$, we have
\begin{equation}
G_+(0)=G_-(0)=G_\star+\epsilon^2\,\delta\mathbf X_-^T\delta\mathbf X_- .
\label{eq:same-gram-initial}
\end{equation}
Uniqueness of the Gram flow~\eqref{eq:gram-gradient-flow} then gives $G_+(\tau)=G_-(\tau)$ for all $\tau$. The two full trajectories differ only by the orientation of the three-dimensional configuration: they are related by the reflection $R={\rm diag}(1,1,-1)$, which is not an $SO(3)$ gauge transformation and reverses all cubic orientation invariants. Direct numerical integration of~\eqref{eq:full-gradient-flow} gives
\begin{equation}
\mathbf X_\star\longrightarrow (\vec p=0,\Sigma=1,\Omega=+1), \qquad
\mathbf X_\star\longrightarrow (\vec p=0,\Sigma=1,\Omega=-1), \label{eq:full-flow-endpoints}
\end{equation}
for the two signs of the negative mode. Thus the two branches of the full unstable manifold end on the two orientation-related physical vacua. In Section~\ref{PT} the perturbative expansion is defined after fixing one orientation sign. For either perturbative sector, exactly one of the two unstable branches therefore connects $X_\star$ to the chosen perturbative vacuum. Reversing that branch gives a unique connecting flow, modulo translation of the flow parameter, from the perturbative critical point to $X_\star$ on the Stokes ray. Consequently
\begin{equation}
|m_{0\star}|=1.\label{eq:magnitude-m-64}
\end{equation}
The overall sign depends on the orientation chosen for $\mathcal J_\star$. We choose the orientation for which this unique connecting trajectory contributes positively, so that
\begin{equation}
m_{0\star}=+1,\qquad \mathcal J_0^+=\mathcal J_0^-+\mathcal J_\star . \label{eq:mplus-64}
\end{equation}
Suppose the perturbative expansion about the physical minimum is
\begin{equation}
\Phi_0(g)=\sum_{n=0}^{\infty} c_n g^n .
\end{equation}
The ordinary Borel transform has a logarithmic singularity at $\zeta=A_\star$ whose coefficient has magnitude
\begin{equation}
|C_{\log}|=\frac{R_{\rm 1loop}}{2\pi}\simeq0.37819516, \label{eq:Clog-64}
\end{equation}
while the Borel--Leroy transform we use has
\begin{equation}
B_{9/2}[\Phi_0](\zeta)={\rm analytic}+C_{9/2}\left(1-\frac{\zeta}{A_\star}\right)^{7/2}+\cdots ,
\end{equation}
with
\begin{equation}
|C_{9/2}|=\frac{R_{\rm 1loop}}{2}\simeq1.18813512 .
\end{equation}
The relation between the ordinary Borel and Borel--Leroy normalizations is derived in Appendix \ref{BorelSing}.

\begin{figure}[htbp]
  \centering
  \includegraphics[width=0.8\textwidth]{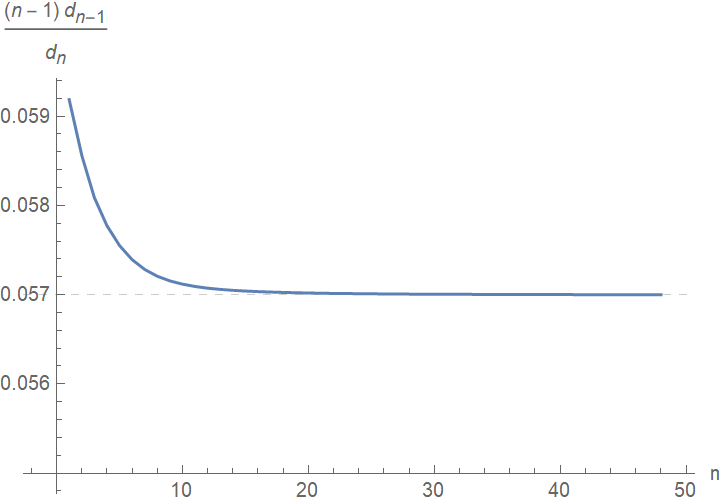}
  \caption{The dashed line shows the value of $a_\star=0.0569957516$. The solid line shows $a_n=(n-1)\frac{d_{n-1}}{d_n}$, with $n=3,4,\cdots,50$.}
  \label{LargeOrders}
\end{figure}

There is an independent check on the location of the leading Borel singularity directly from the perturbative coefficients. Write $d_n=\lambda^n c_n$ and $a_\star=A_\star/\lambda$. If the singularity at $A_\star$ controls the leading large-order behavior, then
\begin{equation}
d_n\sim C_{\log}\,\frac{\Gamma(n)}{a_\star^n}\left(1+O\left(\frac1n\right)\right).
\end{equation}
The simple ratio estimator
\begin{equation}
a_n=(n-1)\frac{d_{n-1}}{d_n}
\end{equation}
constructed from the coefficients displayed in Section 4 gives
\begin{equation}
a_2=0.0557710,\qquad a_3=0.0591862,\qquad a_4=0.0585624,
\end{equation}
already close to
\begin{equation}
a_\star=0.0569957516.
\end{equation}
We extend this to the first 50 orders of perturbation theory in Figure \ref{LargeOrders} and observe a clear convergence to $a_\star$. Thus perturbation theory near the physical vacuum, the independent kernel reduction, and the symmetry-reduced saddle calculation all identify the same nonperturbative scale.

\medskip

After complexification the original integration cycle can be written, away from a Stokes ray, as
\begin{equation}
\mathcal C=\sum_\alpha \nu_\alpha\,\mathcal J_\alpha,\qquad \nu_\alpha=\langle\mathcal C,\mathcal K_\alpha\rangle ,\label{eq:thimble-decomposition-64}
\end{equation}
where $\mathcal K_\alpha$ is the upward-flow cycle dual to $\mathcal J_\alpha$. The integers $\nu_\alpha$ encode which thimbles occur in the chosen exact integration contour. For real positive $g$ we sit on the Stokes ray, so the thimble basis has two lateral limits. Denote the corresponding coefficients of $\mathcal J_\star$ by $\nu_\star^+$ and $\nu_\star^-$. Keeping the physical contour fixed while using \eqref{eq:mplus-64} gives
\begin{equation}
\nu_\star^- - \nu_\star^+=1\label{eq:lateral-incidence-64}
\end{equation}
for the orientation convention chosen above. Equivalently,
\begin{equation}
|\nu_\star^- - \nu_\star^+|=1
\end{equation}
is orientation independent. For a lateral basis in which the additional saddle thimble is absent on one side, the two coefficients are $(0,1)$, or $(1,0)$ if the two lateral directions are interchanged. On the Stokes line itself there is no unique integer which is ``the'' coefficient $\nu_\star$; the unambiguous statement is the unit jump between the two lateral decompositions.

The conclusion is thus sharper than the conclusion we obtained using only local saddle data. The rank-two saddle has potential difference
\begin{equation}
A_\star=0.0569957516133\,\lambda,
\end{equation}
its one-loop normalization is
\begin{equation}
R_{\rm 1loop}=2.37627024,
\end{equation}
and the full-matrix gradient flow shows the Stokes connection coefficient between the perturbative saddle (with a fixed orientation sign) and the rank-two saddle has magnitude one,
\begin{equation}
|m_{0\star}|=1.
\end{equation}
Thus the location, branch exponent and leading coefficient of the associated Borel--Leroy singularity are fixed without assuming unit incidence. The remaining lateral ambiguity concerns the decomposition of the full physical contour on the Stokes ray, not the existence or strength of the Borel singularity itself.

\subsection{A first quantitative test of the bridge equation}
\label{sec:bridge-test}

The analysis of the previous subsection determines the action, one-loop normalization and Stokes connection of the rank-two saddle.  Resurgence predicts more than these leading data: the entire fluctuation series about the rank-two saddle should reappear, order by order, in the subleading large-order behaviour of perturbation theory about the physical minimum.  The purpose of this subsection is to explain this statement explicitly and then test its first nontrivial consequence.

For this discussion it is convenient to remove the inessential scale $\lambda$ and introduce
\begin{equation}
 t={g\over\lambda},\qquad a_\star={A_\star\over\lambda}=0.0569957516133,
 \qquad d_n=\lambda^n c_n .
 \label{eq:t-astar-65}
\end{equation}
Thus the perturbative expansion about the physical minimum may be written as
\begin{equation}
 \Phi_0(t)\equiv\Phi_0(g=\lambda t)=\sum_{n=0}^{\infty}d_n t^n.
\end{equation}

When we cross a Stokes line, the perturbative thimble may jump. This discontinuity is captured by the bridge equation. Let $\Phi_\star^{\rm th}(t)$ denote the fluctuation series carried by the thimble of the rank-two saddle, including its relative one-loop normalization.  Introduce a transseries parameter $\sigma$ and write
\begin{equation}
 \Phi(t,\sigma)=\Phi_0(t)+\sigma e^{-a_\star/t}\Phi_\star^{\rm th}(t)+\cdots .
 \label{eq:transseries-65}
\end{equation}
The parameter $\sigma$ is a bookkeeping device: it records the coefficient multiplying the rank-two saddle sector.  The Picard--Lefschetz result
\begin{equation}
 {\cal J}_0^+={\cal J}_0^-+{\cal J}_\star
\end{equation}
means that crossing the Stokes ray changes this coefficient by one unit, $\sigma\to\sigma+1$.  If $\dot\Delta_{a_\star}$ denotes the pointed alien derivative associated with the Borel singularity at $a_\star$, the infinitesimal form of this translation is the bridge equation
\begin{equation}
 \dot\Delta_{a_\star}\Phi(t,\sigma)={\partial\over\partial\sigma}\Phi(t,\sigma).
\end{equation}
Setting $\sigma=0$ gives
\begin{equation}
 \dot\Delta_{a_\star}\Phi_0(t)=e^{-a_\star/t}\Phi_\star^{\rm th}(t).
\end{equation}
Since the pointed alien derivative is $\dot\Delta_{a_\star}=e^{-a_\star/t}\Delta_{a_\star}$, this may equivalently be written as
\begin{equation}
 \Delta_{a_\star}\Phi_0(t)=\Phi_\star^{\rm th}(t).
\end{equation}
This is the form of the bridge equation that we will use: the Borel singularity of perturbation theory around the physical vacuum carries the fluctuation series around the distant rank-two saddle. It is useful to separate the overall one-loop normalization from the fluctuation series and write
\begin{equation}
 \phi_\star(t)=1+b_1^\star t+b_2^\star t^2+\cdots .
\end{equation}
In this normalization the magnitude of the bridge constant is $R_{\rm 1loop}$.  The overall phase depends on the orientation of the negative mode and on the lateral-resummation convention; for the large-order comparison below only its magnitude is needed.

We can now explain what the bridge equation means in the Borel plane. The dimensionless Borel variable used here is $\xi=\zeta/\lambda$, so that the saddle singularity sits at $\xi=a_\star$. Define the ordinary Borel transform of the perturbative series by
\begin{equation}
 \widehat\Phi_0(\xi)=\sum_{n=0}^{\infty}{d_n\over n!}\,\xi^n .
\end{equation}
The bridge equation says that the singularity of $\widehat\Phi_0$ at $\xi=a_\star$ is not arbitrary: its discontinuity is proportional to the Borel transform of the fluctuation series around the rank-two saddle.  Equivalently, suppressing the convention-dependent overall phase, the local singular part has the form
\begin{align}
\widehat\Phi_0(\xi)\big|_{\rm sing} =-C_{\log}\Bigg[1&+b_1^\star(\xi-a_\star)
+{b_2^\star\over2!}(\xi-a_\star)^2+\cdots\Bigg]\log\left(1-{\xi\over a_\star}\right).\label{eq:singular-germ-65}
\end{align}
The logarithm supplies the branch singularity at $a_\star$, while the analytic function multiplying it is precisely the fluctuation series about the other saddle, written in Borel space. Equation \eqref{eq:singular-germ-65} immediately determines the large-order behaviour of the coefficients $d_n$.  The leading term follows from
\begin{equation}
 -\log(1-x)=\sum_{n=1}^{\infty}{x^n\over n},
\end{equation}
which gives
\begin{equation}
 d_n\sim C_{\log}{\Gamma(n)\over a_\star^n}.
\end{equation}
The first correction follows from
\begin{equation}
 (1-x)\log(1-x)= -x+\sum_{n=2}^{\infty}{x^n\over n(n-1)}.
\end{equation}
Since $\xi-a_\star=-a_\star(1-\xi/a_\star)$, the term proportional to $b_1^\star$ in \eqref{eq:singular-germ-65} contributes
\begin{equation}
 C_{\log}b_1^\star{\Gamma(n-1)\over a_\star^{\,n-1}}.
\end{equation}
More generally, the coefficient $b_k^\star$ produces the $k$th inverse-$n$ correction, so that
\begin{equation}
 d_n\sim C_{\log}\sum_{k\geq0}b_k^\star{\Gamma(n-k)\over a_\star^{\,n-k}}, \qquad b_0^\star=1.
 \label{eq:largeorder-general-65}
\end{equation}
Keeping the first two terms gives the prediction that we test below,
\begin{equation}
 d_n \sim C_{\log}\left[ {\Gamma(n)\over a_\star^n} +b_1^\star{\Gamma(n-1)\over a_\star^{\,n-1}} +\cdots\right]
= C_{\log}{\Gamma(n)\over a_\star^n}\left[1+{b_1^\star a_\star\over n-1}+O\left({1\over n^2}\right)\right].
 \label{eq:largeorder-relative-65}
\end{equation}
The first fluctuation coefficient about the rank-two saddle appears as the first $1/n$ correction to the large-order growth around the physical vacuum.  We now compute $b_1^\star$ independently and test this prediction. The kernel reduction introduced in the previous subsection is convenient for this purpose\footnote{The reduction is only being to evaluate the normalized local fluctuation series. The possible lift/multiplicity ambiguity affects an overall normalization, whereas $b_1^\star$ is a normalized coefficient. The overall one-loop normalization and incidence have already been computed independently in the full variables.}.  Recall that we choose a unit vector $\vec n\in S^3$ and impose $G\vec n=0$.  Then, for fixed $\vec n$ we choose an orthonormal basis $U(\vec n)$ for $\vec n^\perp$ and write $G=UQU^T$, where $Q$ is a symmetric $3\times3$ matrix.  If $\vec q$ denotes the six independent entries of $Q$, the nine primaries are affine-linear functions of $\vec q$,
\begin{equation}
 \vec p=A(\vec n)\vec q-\vec c.
\end{equation}
We then define $K(\vec n)=A(\vec n)^T A(\vec n)$ and complete the square in the six variables $q$
\begin{equation}
 (\vec p\cdot\vec p) = (\vec q-\vec\mu)^T K(\vec n)(\vec q-\vec \mu)+{\cal S}(\vec n),
 \label{eq:complete-square-65}
\end{equation}
where ${\cal S}(\vec n)$ is precisely the reduced function used in the analysis of the previous subsection.  On the complex saddle thimble the six nondegenerate $\vec q$ directions are Gaussian and may be integrated exactly.  Up to an overall constant common to the saddle sectors, the remaining local integral is therefore
\begin{equation}
 {\cal I}(t) \propto (2\pi t)^3 \int_{S^3}{d\Omega_3(\vec n)\over\sqrt{\det K(\vec n)}}\, \exp\left[-{{\cal S}(\vec n)\over2t}\right].
 \label{eq:reduced-integral-65}
\end{equation}
All non-Gaussian fluctuations have been reduced to a three-dimensional saddle expansion.

To perform the local fluctuation expansion about the non-trivial rank-two saddle $\vec n_\star$, let $E_\alpha$, $\alpha=1,2,3$, be an orthonormal basis of the tangent space to $S^3$ at $\vec n_\star$.  A convenient local parametrization is
\begin{equation}
 \vec n(\vec u)= {\vec n_\star+E_\alpha \vec u_\alpha\over\sqrt{1+u^2}},\qquad  u^2=\sum_{\alpha=1}^3u_\alpha^2.
 \label{eq:local-n-65}
\end{equation}
In these coordinates the sphere measure is
\begin{equation}
 d\Omega_3={d^3u\over(1+u^2)^2}.
\end{equation}
It is therefore useful to define
\begin{equation}
 f(\vec u)={1\over2}{\cal S}(\vec n(\vec u)), \qquad {\cal A}(\vec u) = {1\over(1+u^2)^2\sqrt{\det K(\vec n(\vec u))}}.
 \label{eq:f-A-65}
\end{equation}
The saddle is at $\vec u=0$ and $f(\vec 0)=a_\star$. The three eigenvalues of the Hessian of $f$ are approximately
\begin{equation}
 {\rm eig}\,H \simeq \{-0.238897,\;6.41170,\;14.75734\}.
 \label{eq:reduced-hessian-65}
\end{equation}
The single negative eigenvalue is the reduced version of the unstable direction already found in the full Hessian.  Its integration contour is rotated onto the steepest-descent direction.  The resulting overall phase is part of the Stokes constant and does not affect the real normalized fluctuation coefficients computed below.

Expand the reduced action and amplitude as
\begin{align}
f(u)&=a_\star+{1\over2}H_{ij}u_i u_j+{1\over3!}f_{ijk}u_i u_j u_k+{1\over4!}f_{ijkl}u_i u_j u_k u_l+\cdots ,\label{eq:f-expand-65}\\
 {\cal A}(u) &= {\cal A}_0 +{\cal A}_i u_i +{1\over2}{\cal A}_{ij}u_i u_j+\cdots .\label{eq:A-expand-65}
\end{align}
The expansion of the action above \eqref{eq:f-expand-65} defines a propagator which is proportional to $t$, as well as a cubic vertex and a quartic vertex which are both proportional to $t^{-1}$. Rescale $u_i=\sqrt{t}\,z_i$. After rescaling the propagator is independent of $t$, while the ${\cal A}_i$ contribution and the cubic vertex are both $\propto\sqrt{t}$, and the ${\cal A}_{ij}$ contribution and the quartic vertex are $\propto t$. The quadratic part defines the propagator
\begin{equation}
 C_{ij}=(H^{-1})_{ij}.
\end{equation}
All higher corrections are evaluated by Wick contracting the $z_i$.  In particular,
\begin{equation}
 \langle z_i z_j z_k z_l\rangle = C_{ij}C_{kl}+C_{ik}C_{jl}+C_{il}C_{jk},
\end{equation}
while the six-point function is the sum over its fifteen pairings. Keeping all terms of order $t$ gives
\begin{align}
 b_1^\star={1\over{\cal A}_0}\Bigg[ &{1\over2}{\cal A}_{ij} \langle z_i z_j\rangle -{1\over6}{\cal A}_i f_{jkl}\langle z_i z_j z_k z_l\rangle \nonumber\\
 &-{ {\cal A}_0\over24}f_{ijkl} \langle z_i z_j z_k z_l\rangle+{{\cal A}_0\over72}f_{ijk}f_{lmn} \langle z_i z_j z_k z_l z_m z_n\rangle\Bigg].
 \label{eq:b1-wick-65}
\end{align}
The four terms have a simple interpretation: the first comes from the quadratic variation of the measure factor, the second mixes the linear variation of the measure with the cubic interaction, the third is the quartic interaction, and the fourth is the diagram containing two cubic interactions. Thus \eqref{eq:b1-wick-65} is simply the usual two-loop saddle expansion written in the reduced three-dimensional variables.

Evaluating the derivatives of the exact functions in \eqref{eq:f-A-65} at $\vec n_\star$ gives
\begin{equation}
 b_1^\star=2.2289200685,
\end{equation}
and hence
\begin{equation}
 \phi_\star(t)=1+2.2289200685\,t+O(t^2)
 =1+2.2289200685\,{g\over\lambda}+O\!\left({g^2\over\lambda^2}\right).
 \label{eq:phi-star-65}
\end{equation}
There is an independent check of this calculation.  Applying exactly the same reduced saddle expansion to the physical perturbative minimum rather than to $\vec n_\star$ gives
\begin{equation}
 b_1^{(0)}= 7.4921875000= {959\over128}.
 \label{eq:b10-check-65}
\end{equation}
Section 4 gave $c_1={959\over128\lambda}$, or equivalently $d_1=959/128$. Thus the reduced three-dimensional calculation reproduces the first perturbative coefficient obtained independently from the nine-dimensional sheet Jacobian.

Now that all quantities entering the large-order prediction \eqref{eq:largeorder-relative-65} have been determined, we can perform a quantitative test of the Bridge equation. Using 
\begin{equation}
 C_{\log}={R_{\rm1loop}\over2\pi}=0.3781951548\ldots,\qquad b_1^\star a_\star=0.1270389746\ldots,\qquad a_\star= 0.0569957516 \label{eq:largeorder-numbers-65}
\end{equation}
we can evaluate the estimates
\begin{equation}
{\rm leading}-d_n=C_{\log}{\Gamma(n)\over a_\star^n}\qquad
\hat{d}_n=C_{\log}{\Gamma(n)\over a_\star^n}\left[1+{b_1^\star a_\star\over n-1}\right]\label{paramfree}
\end{equation}
There are no fitted parameters in the prediction.  Comparing \eqref{eq:largeorder-relative-65} with the coefficients obtained in Section 4 gives
\begin{equation}
\begin{array}{c|c|c|c}
n& \text{exact }d_n& {\rm leading}-d_n & \hat{d}_n\\ \hline
 2&134.3384094&116.4209102&131.2109033\\
 3&4539.520913&4085.248704&4344.741607\\
 4&232547.8709&215029.1165&224134.8093
\end{array}
\label{eq:bridge-table-65}
\end{equation}
At $n=2$ the relative error improves from about $13.3\%$ to $2.3\%$; at $n=3$ it improves from about $10.0\%$ to $4.3\%$; and at $n=4$ it improves from about $7.5\%$ to $3.6\%$.  These are extremely low orders at which to test an asymptotic large-order formula, so the improvement is already significant. Below we continue this comparison to the first 50 orders of perturbation theory. The last point on the plot is
\begin{equation}
{d_{50}\over\hat{d}_{50}}=1.00017
\end{equation}
demonstrating excellent agreement. Thus, perturbation theory around one sheet knows not merely $A_\star$, but also the one-loop normalization and the first nontrivial fluctuation coefficient of the distant saddle.

\begin{figure}[htbp]
  \centering
  \includegraphics[width=0.75\textwidth]{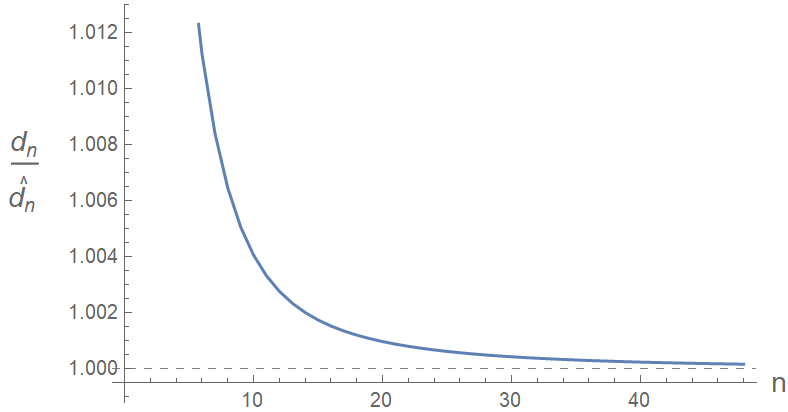}
  \caption{The solid line shows $d_n/\hat{d}_n$, where $\hat{d}_n$ is the parameter-free prediction \eqref{paramfree} including the first fluctuation correction about the rank-two saddle. The dashed line shows the asymptotic value 1.}
  \label{forBidge}
\end{figure}
In this comparison $b_1^\star$ is computed independently from local fluctuations around the distant rank-two saddle.  The bridge equation predicts that this number must reappear in the first inverse-$n$ correction to the large-order behaviour around the physical minimum, and this is what the comparison confirms.  In this sense the perturbative expansion around one Hironaka sheet contains not only the action and one-loop weight of a distant saddle, but also detailed information about the fluctuation theory around that saddle.

The first 50 perturbative coefficients also allow us to probe the Borel plane directly.  Define the truncated ordinary Borel transform
\begin{equation}
\widehat\Phi_0^{(50)}(\xi)=\sum_{n=0}^{50}{d_n\over n!}\,\xi^n.
\label{eq:pade-borel-def}
\end{equation}
We analytically continue this truncated Taylor series using near-diagonal Pad\'e approximants. By a near-diagonal Padé approximant we mean a rational function
\begin{equation}
[L/M](\xi)=\frac{P_L(\xi)}{Q_M(\xi)}
\end{equation}
\begin{figure}[h]
  \centering  \includegraphics[width=0.77\textwidth]{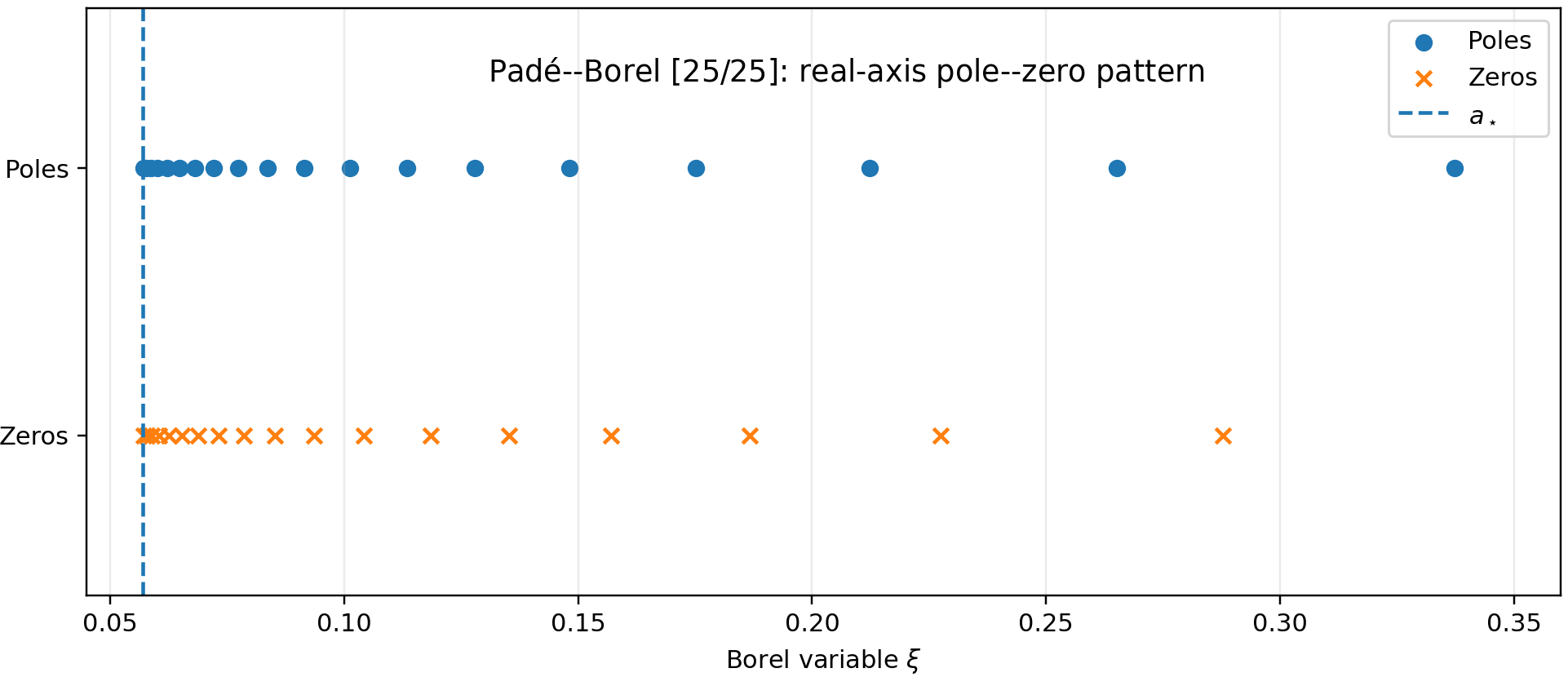}
  \caption{Poles and zeros of the $[25/25]$ Pad\'e approximant to the truncated ordinary Borel transform $\widehat\Phi_0^{(50)}(\xi)$.  Their interlacing condensation along the positive real axis is the characteristic Pad\'e representation of a branch cut beginning at the leading singularity.} \label{fig:pade-borel-cut}
\end{figure}

whose Taylor expansion reproduces $\widehat\Phi_0^{(50)}(\xi)$ up to order $L+M$, with $L$ and $M$ chosen as close as possible, $L\simeq M$. Figure~\ref{fig:pade-borel-cut} shows the poles and zeros of the $[25/25]$ Pad\'e approximant to $\widehat\Phi_0^{(50)}(\xi)$.  They accumulate along the positive real Borel axis beginning near $\xi=a_\star$, which is the characteristic pattern by which a rational Pad\'e approximant represents a branch cut.  Thus the perturbative data themselves reconstruct the expected positive-real Borel cut associated with the rank-two saddle.

\begin{figure}[htbp]
  \centering
  \includegraphics[width=0.82\textwidth]{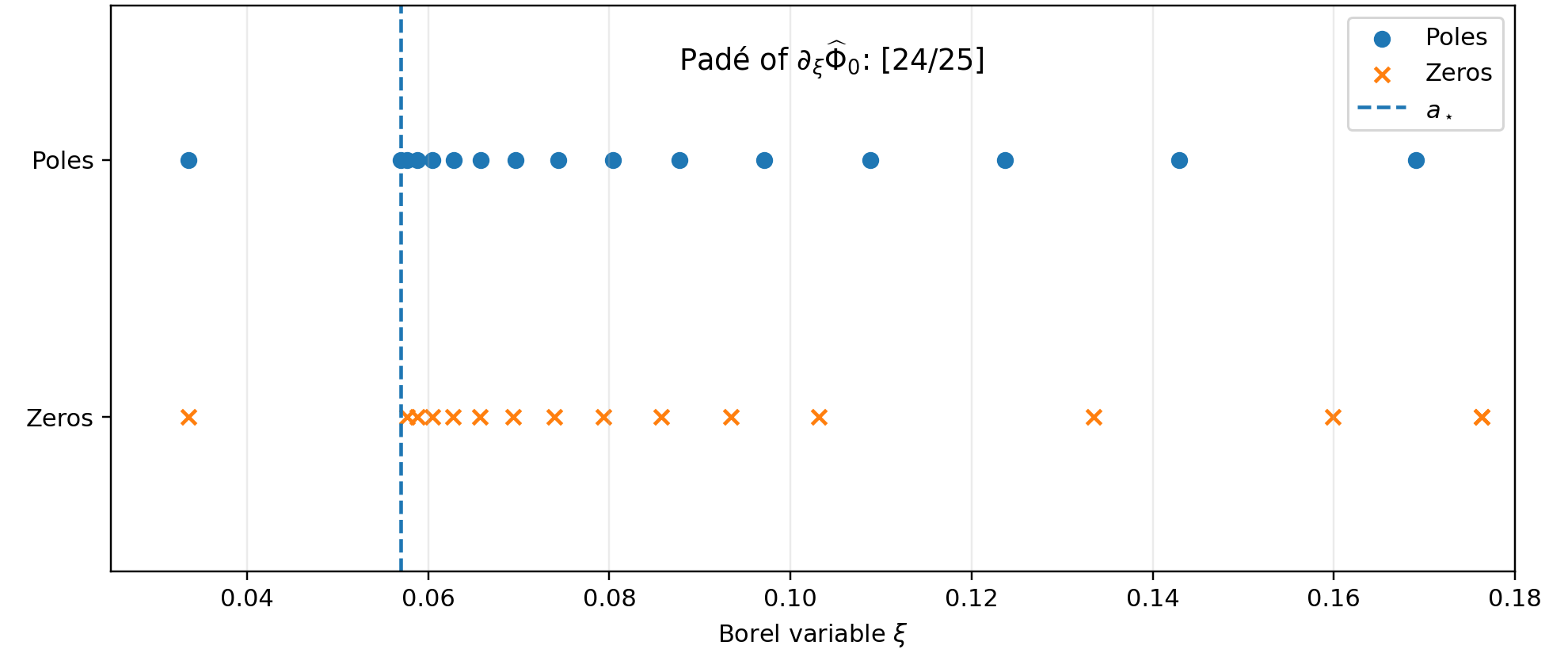}
  \caption{Poles and zeros of the $[24/25]$ Pad\'e approximant to $\partial_\xi\widehat\Phi_0^{(50)}(\xi)$.  Because the leading singularity of $\widehat\Phi_0$ is logarithmic, the derivative has a simple pole at the same location.  The dashed line marks the independently computed value $\xi=a_\star$.}
  \label{fig:pade-borel-deriv}
\end{figure}

In our problem the leading ordinary-Borel singularity is logarithmic, so it is useful to differentiate once with respect to $\xi$.  The log branch point is then converted into a simple pole to leading order.  We therefore also construct a Pad\'e approximant to $\partial_\xi\widehat\Phi_0^{(50)}(\xi)$.  A $[24/25]$ approximant gives
\begin{equation}
\xi_{\rm Pade}=0.0569957965,\qquad a_\star=0.0569957516,\label{eq:pade-a-star}
\end{equation}
in excellent agreement, with a relative difference of about $8\times10^{-7}$.  The same leading pole is stable under small changes of Pad\'e order, for example $[23/26]$, $[25/24]$ and $[26/23]$.  Figure~\ref{fig:pade-borel-deriv} shows the pole-zero pattern for the $[24/25]$ approximant.  This provides a second direct confirmation of the leading Borel singularity: in addition to the saddle calculation and the large-order ratio test, analytic continuation of the perturbative Borel transform itself reconstructs the predicted singularity at $a_\star$.

\section{General Hironaka mechanism}
\label{sec:general_mechanism}

We have developed our arguments using a definite example: four traceless $2\times2$ matrices described by nine primary invariants and an eight-sheeted Hironaka cover. The mechanism we have found, however, does not depend on these details. The essential ingredients are much simpler. At finite $N$, a choice of primary invariants defines a finite projection from the full invariant space to the space coordinatized by the primaries. A generic point in primary space has a finite number of points lying above it, the Hironaka sheets. These sheets can collide. At such a collision the projection ceases to be locally one-to-one, and two things happen at once: the measure obtained after projecting onto the primaries becomes singular, and an action that depends only on the primaries becomes automatically stationary in the directions that are lost by the projection. These are the ingredients that were needed in our explicit calculation, showing that the analysis is more general than the particular example we studied. The purpose of this section is to explain the general mechanism without relying on the special form of the $N=2$, $d=4$ model.

\subsection{Ramification creates new saddle directions}
\label{sec:general_saddles}

Let ${\cal M}$ denote the complexified gauge-invariant configuration space and let
\begin{equation}
    \pi:{\cal M}\longrightarrow {\cal B}
\end{equation}
be the map to the space ${\cal B}$ coordinatized by the primary invariants $p=(p_1,\ldots,p_h)$. The Hironaka decomposition implies that this map is finite: for a generic choice of the primary invariants there are only finitely many points of ${\cal M}$ lying above it. If the number of secondary invariants is $N_s$, then a generic fiber contains $N_s$ points after complexification. Thus, locally and away from collisions, the invariant space looks like $N_s$ separate copies of primary space.

Now suppose that the potential depends only on the primary invariants,
\begin{equation}
    V_{\rm lift}(x)=V\bigl(\pi(x)\bigr),\qquad x\in {\cal M}.
\end{equation}
This is exactly the situation we consider in our analysis. Varying the action gives
\begin{equation}
    d(V_{\rm lift})_x=(d\pi_x)^T dV_{\pi(x)}.
\end{equation}
At a generic point of a sheet, the map $\pi$ is locally invertible. Every infinitesimal displacement in invariant space changes the primaries to first order. In that case
\begin{equation}
    d(V_{\rm lift})_x=0    \qquad\Longleftrightarrow\qquad    dV_{\pi(x)}=0.
\end{equation}
Thus an ordinary, unramified point cannot generate a new saddle: critical points upstairs are simply the lifts of critical points of the potential in primary space.

The situation changes at a sheet collision. At a collision the differential $d\pi_x$ loses rank. Consequently there are nonzero directions in ${\cal M}$ along which the primaries do not change to first order. Since the action is a function only of the primaries, its first derivative vanishes automatically in these directions. This is the general reason that ramification can produce additional saddles even at points for which $dV\neq0$ in the full primary space.

The simplest local model makes this completely transparent. Near a generic collision of two sheets one can choose local coordinates such that
\begin{equation}
    p_1=u^2,\qquad p_a=v_a,\qquad a=2,\ldots,h.
\end{equation}
The two sheets correspond to $u=\pm\sqrt{p_1}$ and meet at $u=0$. The action becomes
\begin{equation}
    V_{\rm lift}(u,v)=V(u^2,v),
\end{equation}
and hence
\begin{equation}
    \frac{\partial V_{\rm lift}}{\partial u}
    =2u\frac{\partial V}{\partial p_1}.
\end{equation}
At the collision, $u=0$, this derivative vanishes irrespective of the value of $\partial V/\partial p_1$. The ramified direction is therefore automatically stationary. The remaining saddle equations are simply
\begin{equation}
    \frac{\partial V}{\partial p_a}=0,
    \qquad a=2,\ldots,h,
\end{equation}
which say that $V$ must be stationary along the branch locus itself. Thus a sheet collision removes one or more first-order directions from the primary description. For an action depending only on the primaries, stationarity in those directions is automatic. We must then impose stationarity only along the collision locus itself. At a smooth ramification point this gives a genuine critical point of the lifted action. At a more singular collision it gives a candidate saddle\footnote{At a more singular collision the invariant-space locus is itself singular, so the reduced stationarity conditions need not capture every variation of the original matrix variables.} and we must check the result directly in the original matrix variables.

In the example of Section~6 the relevant Gram matrix has rank two. There are two physical directions transverse to the rank-two locus that become invisible to the Gram invariants at first order, while the remaining seven directions are tangent to the rank-two locus. The calculation performed there is therefore the higher-corank version of the simple local model above; the full matrix saddle equation provides the additional direct confirmation.

\subsection{Why the projected measure becomes singular}
\label{sec:general_jacobian}

The loss of local invertibility explains the singularity of the sheet Jacobian. To see this in a form that applies beyond our example, introduce local secondary coordinates $\vec{y}=(y_1,\ldots,y_m)$ and suppose that the invariant space is described locally by $m$ equations
\begin{equation}
    F_A(\vec{p},\vec{y})=0,    \qquad A=1,\ldots,m.
\end{equation}
Away from a sheet collision these equations can be solved for the secondary coordinates, $\vec{y}=\vec{y}_r(\vec{p})$, where $r$ labels different sheets. The condition for this to be possible is
\begin{equation}
    \det\left(\frac{\partial F_A}{\partial y_B}\right)\neq0.
\end{equation}
Consider now an invariant-space integral of the form
\begin{equation}
    Z=\int d^h p\,d^m y\; \prod_{A=1}^m\delta\left(F_A(\vec{p},\vec{y})\right) \mu(\vec p,\vec y)\,e^{-V(\vec p)/g}.
\end{equation}
Integrating over the secondary coordinates gives a sum over sheets,
\begin{equation}
    Z=\sum_r\int d^h p\; J_r(\vec{p})e^{-V(\vec{p})/g},
\end{equation}
with
\begin{equation}
    J_r(\vec{p}) =\frac{\mu\bigl(p,y_r(p)\bigr)}{\left|\det\left(\partial F_A/\partial y_B\right)_{\vec{y}=\vec{y}_r(p)}\right|}
\end{equation}
on a real integration cycle. After complexification the absolute value is replaced by the corresponding holomorphic residue. The denominator of the projected measure is precisely the Jacobian that tests whether the secondary coordinates can be solved as functions of the primaries. At a sheet collision this determinant vanishes. Unless the numerator vanishes at the same rate, the contribution from an individual sheet becomes singular. In the one-secondary case the statement reduces to
\begin{equation}
    F(p,y_r(p))=0,    \qquad\qquad    J_r(\vec{p})\propto \frac{1}{\partial_yF(\vec{p},\vec{y}_r(\vec{p}))}.
\end{equation}
Two roots collide when
\begin{equation}
    F(p_\star,y_\star)=0, \qquad\qquad \partial_y F(p_\star,y_\star)=0,
\end{equation}
and hence the branchwise Jacobian is singular there. This is exactly the structure of our example, where $y=\Sigma$ and the perturbative measure contains $1/\partial_\Sigma F$. The singularity appears because primary coordinates cease to be good local coordinates at the collision and nothing is necessarily singular in the original matrix variables. In the simple two-sheet model $s=u^2$, the measure $du$ is perfectly regular at $u=0$. Expressed in terms of the projected coordinate $s$, however,
\begin{equation}
    du=\frac{ds}{2u} =\pm\frac{ds}{2\sqrt{s}}.
\end{equation}
The contribution from either sheet behaves as
\begin{equation}
    J_\pm(s)\sim s^{-1/2}.
\end{equation}
The divergence is the familiar Jacobian singularity produced when two branches of a coordinate transformation meet. The same argument gives the local behavior for a higher ramification. If $e$ sheets meet with local form $s=u^e$, then
\begin{equation}
    du=\frac{1}{e}s^{1/e-1}ds,\qquad {\rm and\,\, hence}\qquad    J(s)\sim s^{-(e-1)/e}.
\end{equation}
The square-root singularity of the simple two-sheet collision is the universal local behavior of a simple ramification.

There are two qualifications. First, the statement is about the density obtained after projection to the primary variables, not about a singularity of the original matrix integral. Second, a zero in the numerator, or a cancellation after summing several sheets, can soften or remove the singularity. In the absence of such a cancellation, however, a collision of Hironaka sheets produces a singular branchwise density in primary space.

\subsection{The discriminant is the natural global object}\label{sec:general_discriminant}

It is useful to have a way of locating sheet collisions without choosing a particular labeling of the sheets. This can be achieved using the discriminant. Consider first a patch described by a single secondary $y$ satisfying
\begin{equation}
    F(\vec{p},y)=0.
\end{equation}
For fixed $\vec{p}$, let the roots be $y_1(p),\ldots,y_r(p)$. The discriminant is, up to an overall factor,
\begin{equation}
    {\rm Disc}_y F \propto \prod_{\alpha<\beta} \bigl(y_\alpha(p)-y_\beta(p)\bigr)^2.
\end{equation}
It vanishes precisely when two or more sheets collide. Equivalently,
\begin{equation}
    {\rm Disc}_y F =0 \qquad\Longleftrightarrow\qquad F=\partial_y F=0
\end{equation}
for some value of $y$. The same derivative $\partial_y F$ that detects the collision also appears in the denominator of the sheet Jacobian. In the $\Sigma$ description of our example, this relation is direct
\begin{equation}
\text{sheet collision}\quad\Longleftrightarrow\quad\partial_\Sigma F=0\quad\Longrightarrow\quad
    J\sim\frac{1}{\partial_\Sigma F}\;\text{singular}.
\end{equation}
This formulation also makes it clear why the physics cannot depend on a particular choice of secondary invariants. The secondary invariants provide a basis for the finite algebra over the primary invariants, but they do not canonically label the individual sheets. What is invariant is the locus where the finite projection degenerates. We can give one convenient basis-independent description of the same locus: Let $\cR$ be the invariant ring and let
\begin{equation}
    \cP=\mathbb C[p_1,\ldots,p_h]
\end{equation}
be the polynomial ring generated by the primaries. For any module basis $\{\eta_A\}$ of $\cR$ over $\cP$, form the trace-pairing matrix
\begin{equation}
    \mathsf T_{AB}(p)=\operatorname{Tr}_{\cR/\cP}\!\left(\eta_A\eta_B\right).
\end{equation}
Its determinant is the discriminant of the finite algebra, up to the expected nonzero factor associated with a change of basis. Thus the vanishing locus of $\det\mathsf T(p)$ is intrinsic to the chosen projection onto primary space. In algebraic language this is the ramification/discriminant structure of the finite extension; physically it is simply the locus in primary space at which distinct invariant configurations can no longer be separated by the primaries.

\subsection{From ramification to large-order behavior}\label{sec:general_resurgence}

We can now return to resurgence. In our explicit calculation, perturbation theory about the physical sheet was generated entirely by derivatives of the sheet Jacobian. Schematically,
\begin{equation}
    c_n\sim \frac{1}{n!}\,\Delta^n J_+(p)\big|_{p=0},
\end{equation}
with the precise normalization given in Section~4. High derivatives about the origin are sensitive to the nearest singularities of the analytically continued function $J_+(p)$. The discussion above shows that the natural singularities of this function occur where the finite Hironaka projection ramifies. Ramification is not enough to identify a Borel singularity associated with a new saddle. The collision point must also satisfy the remaining saddle equations. Indeed, for an action depending only on the primaries, $S=V(p)$ is  automatically stationarity in the ramified directions, but we must still require that
\begin{equation}
 d\!\left(V\big|_{\text{collision locus}}\right)=0
 \end{equation}
to obtain a ramification saddle, at the point $p_\star$. The difference in the action
\begin{equation}
A_\star=V(p_\star)-V(p_0)
\end{equation}
is then  a candidate Borel action. The reason why we stress that this is a ``candidate'' is that we still need to check \emph{if the corresponding thimbles are connected by the appropriate Picard--Lefschetz flow}. That is a global dynamical question, not a consequence of invariant theory alone. In Section~\ref{Ram} we checked this additional condition explicitly for the rank-two saddle and found a unit Stokes connection. The Hironaka cover identifies where non-perturbative saddles can naturally arise; Picard--Lefschetz theory determines which of them actually communicate with the perturbative sector under consideration.

This separation of roles is natural: the Hironaka decomposition is kinematical. It is fixed by the finite-$N$ invariant ring before an action is chosen. Its discriminant records the places where the finite collection of sheets comes together. Once an action depending on the primaries is specified, these same loci acquire dynamical significance because the ramified directions are automatically stationary. Resurgence then detects the saddles that are globally connected to the perturbative vacuum through the resulting Borel singularities.

The $N=2$, $d=4$ model therefore provides more than an isolated example. It gives an explicit realization of a general connection between finite-$N$ invariant theory and non-perturbative saddle structure: the finite Hironaka cover organizes the possible sheets, its ramification locates their collisions, the projected measure becomes singular at those collisions, and a primary-only action can turn suitable collision points into genuine saddles whose actions govern the large-order behavior of perturbation theory.

\section{Discussion}

The main result of this work is that finite-$N$ invariant theory can provide concrete information about the non-perturbative structure detected by resurgence. In the example studied here, the Hironaka decomposition determines a finite branched cover of primary-invariant space, and the ramification of this cover identifies loci at which the primary description degenerates. We found that one such degeneration is associated with a genuine rank-two saddle of the original matrix integral. Its action, one-loop normalization and Picard--Lefschetz connection determine the leading Borel singularity of perturbation theory, as independently confirmed by the large-order data, about the physical vacuum. Going beyond the leading exponential scale, the first fluctuation correction about the distant saddle reproduces the first subleading large-order correction to perturbation theory, with no fitted parameters. Thus the relation between invariant theory and resurgence is visible not only in the location of a Borel singularity, but also in the detailed bridge between perturbative expansions around distinct saddles.

It is useful to emphasize the different roles played by invariant theory and dynamics. The Hironaka decomposition is kinematical: it is fixed once the finite-$N$ invariant ring and a choice of primary invariants are specified. It determines the finite cover and its discriminant, and hence identifies the loci where distinct invariant configurations become indistinguishable when described only by the primaries. The action then supplies the dynamical input. For an action of the form
\begin{equation}
V=V(p_1,\ldots,p_h),
\end{equation}
the loss of rank of the projection at a ramification point makes the action automatically stationary in the ramified directions. One must still impose stationarity along the collision locus itself, and even then the resulting critical point contributes to the large-order behavior of a chosen perturbative sector only if the corresponding thimbles are connected. In this sense, invariant theory tells us where non-perturbative saddles can naturally occur, the action determines which points on these loci are genuine saddles and Picard--Lefschetz theory determines which of those saddles are resurgently connected to the perturbative vacuum.

The assumption that the potential depends only on the primary invariants deserves a comment. At the small value $N=2$ used in our explicit example, secondary structure already appears at low degree, so this is a genuine restriction on the choice of dynamics. The situation is different in the large-$N$ regime that motivates the holographic interpretation. Let $\Delta_V$ denote the maximal degree of the interaction terms and let $\Delta_{\rm oc}(N)$ denote the overcrowding scale, namely the degree at which genuinely finite-$N$ relations become unavoidable~\cite{deMelloKoch:2026pck}. For a degree-bounded interaction, $\Delta_V$ is held fixed as $N$ is increased, whereas $\Delta_{\rm oc}(N)$ grows with $N$. Consequently,
\begin{equation}
\frac{\Delta_V}{\Delta_{\rm oc}(N)}\longrightarrow 0 \qquad (N\to\infty).
\end{equation}
For sufficiently large $N$, a fixed-degree potential therefore lies parametrically below the overcrowding scale. In this regime it is natural to regard the ordinary local dynamics as living entirely in the primary sector, with the finite-$N$ secondary structure entering through the global geometry of invariant space rather than through explicit secondary dependence in the local potential. From this point of view, primary-only potentials are not an unnatural fine tuning of the large-$N$ dynamics, but a natural class of bounded-degree interactions for studying how genuinely finite-$N$ global structure becomes visible non-perturbatively.

There is a complementary large-$N$ perspective which is obscured by our choice $N=2$. Once the invariant variables are normalized so that they remain finite at large $N$, the invariant-space effective action carries the usual overall large-$N$ factor. Schematically, $Z\sim \int[d\phi ]e^{-N^2 S_{\rm eff}[\phi]}$, so that fluctuations about a saddle are organized as a $1/N$ loop expansion. A second saddle associated with the ramification geometry would then contribute with a weight of the form
\begin{equation}
\exp\{-N^2[S_{\rm eff}(\phi_\star)-S_{\rm eff}(\phi_0)]\},
\end{equation}
up to the detailed $N$-scaling appropriate to the model. Thus the finite-cover mechanism isolated here is naturally situated in the same invariant-variable framework in which the large-$N$ expansion is performed. Our choice $N=2$ makes this scaling invisible, but allows the underlying finite-$N$ geometry and its resurgent consequences to be studied explicitly.

Finally, the Hironaka cover is fixed entirely by the finite-$N$ trace relations and is therefore an intrinsically finite-$N$ structure. It is remarkable that its geometry is reflected so directly in the resurgent structure of the theory, with ramification of the cover organizing the non-perturbative saddles that control the large-order behavior of perturbation theory.

\begin{center} 
{\bf Acknowledgements}
\end{center}
We are deeply grateful to Jie Gu for his exceptionally patient and illuminating explanations of resurgence. This work was supported by a start up research fund of Huzhou Normal University, a Zhejiang Province talent award and by a Changjiang Scholar award.

\appendix

\section{Finding the stationary point}\label{statpoint}

The symmetry reduced ansatz is
\begin{equation}
p_1=p_2=x,\qquad p_3=w,\qquad p_4=v,\qquad p_5=p_6=0, \qquad p_7=z,\qquad p_8=p_9=y.
\end{equation}
On this subspace we have $F=\frac1{16}P_+P_-$ where
\begin{equation}
P_+=-4(\Sigma+y)^2+(1+x+z)(12+3v+4w),
\end{equation}
\begin{equation}
P_-=-4(\Sigma-y)^2 +(1+x-z)(4+v-4w).
\end{equation}
The potential becomes
\begin{equation}
\frac V\lambda=x^2+\frac12w^2+\frac12v^2+\frac12z^2+y^2.
\end{equation}
To simplify the notation, define the new expressions
\begin{equation}
U=1+x+z,\qquad L=1+x-z,\qquad C_+=12+3v+4w,\qquad C_-=4+v-4w.
\end{equation}
Then
\begin{equation}
P_+=-4(\Sigma+y)^2+UC_+,\qquad P_-=-4(\Sigma-y)^2+LC_-.
\end{equation}
The rank-two collision is obtained by imposing $P_+=0$ and $P_-=0$. We begin by eliminating $y$ and $\Sigma$: the two constraints say
\begin{equation}
4(\Sigma+y)^2=UC_+,\qquad 4(\Sigma-y)^2=LC_-.
\end{equation}
For the real branch containing the saddle we quoted, both $\Sigma+y$ and $\Sigma-y$ are positive, so
\begin{equation}
\Sigma+y=\frac12\sqrt{UC_+},\qquad \Sigma-y=\frac12\sqrt{LC_-}.
\end{equation}
Adding and subtracting these equations gives
\begin{equation}
\Sigma= \frac14\left(\sqrt{UC_+}+\sqrt{LC_-}\right)=\frac{R+T}{4}\qquad {\rm and}\qquad y=\frac14\left(\sqrt{UC_+}-\sqrt{LC_-}\right)=\frac{R-T}{4},
\end{equation}
where $R=\sqrt{UC_+}$ and $T=\sqrt{LC_-}$. Consequently, once $x,w,v,z$ have been found, $y$ and $\Sigma$ are determined. We now reduce the constrained problem to four variables. The potential restricted to the collision locus becomes
\begin{equation}
f(x,w,v,z)\equiv\frac{V_{\rm restricted}}{\lambda}=x^2+\frac12w^2+\frac12v^2+\frac12z^2
+\frac1{16}(R-T)^2,
\end{equation}
with
\begin{equation}
R=\sqrt{(1+x+z)(12+3v+4w)},\qquad T=\sqrt{(1+x-z)(4+v-4w)}.
\end{equation}
Now we simply require stationarity of the restricted potential
\begin{equation}
\frac{\partial f}{\partial x}=\frac{\partial f}{\partial w}=\frac{\partial f}{\partial v}=\frac{\partial f}{\partial z}=0.
\end{equation}
These four equations determine $x,w,v,z$. Let's write the four equations explicitly. Using $R=\sqrt{UC_+}$ and $T=\sqrt{LC_-}$, we have
\begin{equation}
\frac{\partial R}{\partial x}=\frac{C_+}{2R},\qquad\frac{\partial T}{\partial x}=\frac{C_-}{2T}\qquad\Rightarrow\qquad
0=2x+\frac{R-T}{16}\left(\frac{C_+}{R}-\frac{C_-}{T}\right).
\end{equation}
Similarly,
\begin{equation}
\frac{\partial R}{\partial w}=\frac{2U}{R},\qquad\frac{\partial T}{\partial w}=-\frac{2L}{T},\qquad\Rightarrow\qquad
0=w+\frac{R-T}{4}\left(\frac{U}{R}+\frac{L}{T}\right).
\end{equation}
For $v$,
\begin{equation}
\frac{\partial R}{\partial v}=\frac{3U}{2R},\qquad\frac{\partial T}{\partial v}=\frac{L}{2T},\qquad\Rightarrow\qquad
0=v+\frac{R-T}{16}\left(\frac{3U}{R}-\frac{L}{T}\right).
\end{equation}
Finally,
\begin{equation}
\frac{\partial R}{\partial z}=\frac{C_+}{2R},\qquad\frac{\partial T}{\partial z}=-\frac{C_-}{2T},\qquad\Rightarrow\qquad
0=z+\frac{R-T}{16}\left(\frac{C_+}{R}+\frac{C_-}{T}\right).
\end{equation}
Thus the entire calculation boils down to solving the above four nonlinear equations.  This gives the lowest positive-action solution we find in this symmetry sector. A Newton solver or Mathematica \texttt{FindRoot} gives the nontrivial solution
\begin{equation}
x=-0.0384062783210153,\qquad w=-0.117388935820739,
\end{equation}
\begin{equation}
v=-0.00929473664615217,\qquad z=-0.224833449493010.
\end{equation}
$y$ and $\Sigma_\star$ follow algebraically. We find $y=0.152683961836736\ldots$ and $\Sigma_\star=1.30287686400498\ldots$. Finally, substitution into the potential gives
\begin{equation}
\frac{V_\star}{\lambda}=x^2+\frac12w^2+\frac12v^2+\frac12z^2+y^2=0.0569957516133196\ldots,
\end{equation}

\paragraph{Summary:} $P_+=P_-=0$ first determines the ramification/collision surface; solving those constraints eliminates $y,\Sigma$; stationarity of $V$ on that surface gives four equations for $x,w,v,z$; and then $y,\Sigma$ are reconstructed algebraically.

\section{The Metric on $SO(3)$ Orbits}\label{OrbMet}

Let an infinitesimal rotation be parametrized by a three-vector $\vec{\omega}=(\omega_1,\omega_2,\omega_3)$. Under this rotation, $\delta \vec x_a=\vec{\omega}\times\vec x_a$. The natural metric on the original configuration space is
\begin{equation}
ds^2=\sum_{a=1}^4 d\vec x_a\cdot d\vec x_a.
\end{equation}
Restricting this metric to infinitesimal rotations, i.e. restricting to $SO(3)$ orbits, gives
\begin{equation}
ds_{\rm orb}^2=\sum_{a=1}^4|\vec{\omega}\times\vec x_a|^2.
\end{equation}
Now use the elementary identity $|\vec{\omega}\times\vec x|^2=|\vec{\omega}|^2|\vec x|^2
-(\vec{\omega}\cdot\vec x)^2$ to obtain
\begin{equation}
ds_{\rm orb}^2=|\vec{\omega}|^2\sum_{a=1}^4|\vec x_a|^2-\sum_{a=1}^4
(\vec{\omega}\cdot\vec x_a)^2.
\end{equation}
The two terms can be written in matrix notation. In terms of $\mathbf X=(\vec x_1,\vec x_2,\vec x_3,\vec x_4)$, we have
\begin{equation}
\mathbf X\mathbf X^T=\sum_{a=1}^4\vec x_a\,\vec x_a^{\,T}\qquad\Rightarrow\qquad {\rm Tr}(\mathbf X\mathbf X^T)=\sum_{a=1}^4|\vec x_a|^2.
\end{equation}
Also,
\begin{equation}
\sum_{a=1}^4(\vec{\omega}\cdot\vec x_a)^2=\sum_a\vec{\omega}^T\vec x_a\vec x_a^{\,T}\vec{\omega}=\vec{\omega}^T\mathbf X\mathbf X^T\vec{\omega}.
\end{equation}
Thus
\begin{equation}
ds_{\rm orb}^2=\vec{\omega}^T\left[\bigl({\rm Tr}\mathbf X\mathbf X^T\bigr)I_3-\mathbf X\mathbf X^T\right]\vec{\omega}.
\end{equation}
By definition, if $ds_{\rm orb}^2=\vec{\omega}^T M_{\rm orb}\vec{\omega}$, then
\begin{equation}
M_{\rm orb}=\bigl({\rm Tr}\mathbf X\mathbf X^T\bigr)I_3-\mathbf X\mathbf X^T.
\end{equation}
This matrix is the \emph{moment-of-inertia tensor} of four unit point masses located at the positions $\vec x_a$
\begin{equation}
I_{ij}=\sum_a\left(|\vec x_a|^2\delta_{ij}-x_{a,i}x_{a,j}\right).
\end{equation}
This is of course expected, since both calculations measure \emph{how much the configuration moves when we rotate it about a given axis.}

Finally let us recall why $\det M_{\rm orb}$ enters the one-loop factor. The three rotational zero modes are parametrized by $\omega_1,\omega_2,\omega_3$. Their induced volume element is therefore
\begin{equation}
d\mu_{\rm orb}\propto\sqrt{\det M_{\rm orb}}\, d^3\omega.
\end{equation}
When comparing the saddle with the perturbative minimum, the geometry of the rotational orbit is different at the two configurations. Hence their collective-coordinate measures differ by
\begin{equation}
\sqrt{\frac{\det M_{{\rm orb},\star}}{\det M_{{\rm orb},0}}}.
\end{equation}

\section{Borel vs. Borel-Leroy Singularity}\label{BorelSing}

Our goal in this Appendix is to explain and distinguish two statements. The ordinary Borel transform of $\Phi_0$ has a logarithmic singularity at $\zeta=A_\star$ of the form $\log\left(1-\frac{\zeta}{A_\star}\right)$ with logarithmic coefficient of magnitude
\begin{equation}
\frac{R_{\rm 1loop}}{2\pi}\label{Bcoeff}
\end{equation}
for unit incidence. The Borel--Leroy transform considered here, $B_{9/2}[\Phi_0](\zeta)$, instead has an algebraic singularity of the form $\left(1-\frac{\zeta}{A_\star}\right)^{7/2}$ whose coefficient has magnitude
\begin{equation}
\frac{R_{\rm 1loop}}{2}.\label{BLcoeff}
\end{equation}
A key take away message is therefore that the Borel and the Borel-Leroy transforms have singularities at the same locations.

Suppose $X_0$ is the perturbative minimum and $X_\star$ is the neighboring saddle, with
\begin{equation}
A_\star=V_\star-V_0>0.
\end{equation}
Across the corresponding Stokes ray, the thimble decomposition changes. If the incidence is one, then schematically
\begin{equation}
\mathcal J_0^+=\mathcal J_0^-+m_{0\star}\mathcal J_\star,\qquad m_{0\star}=\pm1.
\end{equation}
Consequently the two lateral resummations of the perturbative series differ by the contribution of the saddle
\begin{equation}
{\rm Disc}\Phi_0(g)\equiv\St_+\Phi_0(g)-\St_-\Phi_0(g)\propto m_{0\star} e^{-A_\star/g}\Phi_\star(g).
\end{equation}
Including the one-loop normalization,
\begin{equation}
{\rm Disc}\Phi_0(g)\sim m_{0\star} R_{\rm 1loop}e^{-A_\star/g}\left(1+O(g)\right),\label{discont}
\end{equation}
up to the conventional phase associated with the negative mode.

What Borel singularity gives precisely this discontinuity? Take the ordinary Borel transform of
\begin{equation}
\Phi_0(g)=\sum_{n=0}^\infty c_n g^n,\qquad {\rm given\,\,by}\qquad \widehat\Phi_0(\zeta)=\sum_{n=0}^\infty
\frac{c_n}{n!}\zeta^n.
\end{equation}
Suppose near $\zeta=A_\star$ it behaves as
\begin{equation}
\widehat\Phi_0(A_\star+t)\sim C\,\widehat\Phi_\star(t)\log t+\text{regular}.
\end{equation}
Using the discontinuity of the logarithm ${\rm Disc}\log t=2\pi i$, we have
\begin{equation}
{\rm Disc}\widehat\Phi_0(A_\star+t)=2\pi i\,C\,\widehat\Phi_\star(t).
\end{equation}
Inserting this into the Laplace integral, shifting $\zeta=A_\star+t$ and pulling out $e^{-A_\star/g}$, one obtains
\begin{equation}
{\rm Disc}\Phi_0(g)=2\pi i\,C\,e^{-A_\star/g}\Phi_\star(g).
\end{equation}
Comparing to \eqref{discont}, up to orientation and phase conventions, we have
\begin{equation}
C=\frac{m_{0\star} R_{\rm 1loop}}{2\pi i}.
\end{equation}
Taking the magnitude for $|m_{0\star}|=1$, we obtain
\begin{equation}
|C|=\frac{R_{\rm 1loop}}{2\pi},
\end{equation}
which proves \eqref{Bcoeff}.

There is an even simpler derivation from the large-order coefficients. A saddle at $A_\star$ with the same Gaussian $g$-power\footnote{By ``the same Gaussian $g$-power'' we mean that, after factoring out $e^{-S_\sigma/g}$, the quadratic fluctuation integrals about the two saddles carry the same overall factor $g^\alpha$ (here $g^{9/2}$), so their ratio introduces no additional power of $g$.} as the original saddle produces the behaviour
\begin{equation}
c_n\sim C\,\frac{\Gamma(n)}{A_\star^n}\qquad (n\to\infty),
\end{equation}
where $C$ contains the Stokes/incidence and one-loop information. The ordinary Borel transform therefore behaves as
\begin{equation}
\widehat\Phi_0(\zeta)\sim C\sum_{n=1}^\infty\frac{\Gamma(n)}{\Gamma(n+1)}\left(\frac{\zeta}{A_\star}\right)^n=
C\sum_{n=1}^\infty \frac1n \left(\frac{\zeta}{A_\star}\right)^n.
\end{equation}
Using
\begin{equation}
\sum_{n=1}^{\infty}\frac{x^n}{n}=-\log(1-x),
\end{equation}
we get
\begin{equation}
\widehat\Phi_0(\zeta)\sim-C\log\left(1-\frac{\zeta}{A_\star}\right).
\end{equation}
So the logarithm is simply the Borel-plane representation of the factorial growth $c_n\sim \Gamma(n)A_\star^{-n}$.

Now let's extend the analysis to the Borel--Leroy transform we are using. The exact Borel--Leroy transform we use is
\begin{equation}
B_{9/2}[\Phi_0](\zeta)=\sum_{n=0}^{\infty} c_n\,\frac{\Gamma(9/2)}{\Gamma(n+9/2)}\zeta^n.
\end{equation}
Now insert the saddle large-order behaviour
\begin{equation}
c_n\sim C\,\frac{\Gamma(n)}{A_\star^n}\qquad\Rightarrow\qquad B_{9/2}[\Phi_0](\zeta)\sim C\Gamma(9/2)\sum_{n=1}^{\infty}\frac{\Gamma(n)}{\Gamma(n+9/2)} \left(\frac{\zeta}{A_\star}\right)^n.
\end{equation}
For large $n$,
\begin{equation}
\frac{\Gamma(n)}{\Gamma(n+9/2)}\sim n^{-9/2}.
\end{equation}
A power series whose coefficients behave as $n^{-9/2}$ has, near its radius of convergence, a nonanalytic part behaving as
\begin{equation}
\left(1-\frac{\zeta}{A_\star}\right)^{7/2}.
\end{equation}
Thus the singularity of the $B_{9/2}$ transform must be of the form
\begin{equation}
B_{9/2}[\Phi_0](\zeta)=\text{analytic}+C_{9/2}\left(1-\frac{\zeta}{A_\star}\right)^{7/2}+\cdots .
\end{equation}
To determine its coefficient, use
\begin{equation}
[x^n](1-x)^{7/2}\sim\frac{n^{-9/2}}{\Gamma(-7/2)}.
\end{equation}
Matching this with the coefficients in the preceding equation gives
\begin{equation}
C_{9/2}=C\,\Gamma(9/2)\Gamma(-7/2)=\pi C.
\end{equation}
Thus for unit incidence
\begin{equation}
|C_{9/2}|=\frac{R_{\rm 1loop}}{2},
\end{equation}
proving \eqref{BLcoeff}. The singularity is therefore not a bare logarithm. It has a branch cut, because $7/2$ is noninteger, but it is an \emph{algebraic branch point}, not a logarithmic branch point.

Why did the logarithm disappear? Changing from the ordinary Borel transform to a Borel--Leroy transform changes the local form of the singularity. For the ordinary Borel transform,
\begin{equation}
\frac{\Gamma(n)}{\Gamma(n+1)}=\frac1n
\end{equation}
and therefore
\begin{equation}
\sum\frac{x^n}{n}=-\log(1-x).
\end{equation}
For the Borel--Leroy transform,
\begin{equation}
\frac{\Gamma(n)}{\Gamma(n+9/2)}\sim n^{-9/2},
\end{equation}
and therefore the singularity becomes $(1-x)^{7/2}$. The location $\zeta=A_\star$ is unchanged, and the same saddle/Stokes data determine the strength.

\end{document}